\documentclass[12pt]{article}

\def\hybrid{\topmargin 0pt      \oddsidemargin 0pt
        \headheight 0pt \headsep 0pt
        \voffset=-0.5cm
        \textwidth 6.5in        % US paper
        \textheight 9in         % US paper
        \marginparwidth 0.0in
        \parskip 5pt plus 1pt   \jot = 1.5ex}
\catcode`\@=11
\def\marginnote#1{}

\newcount\hour
\newcount\minute
\newtoks\amorpm
\hour=\time\divide\hour by60
\minute=\time{\multiply\hour by60 \global\advance\minute by-\hour}
\edef\standardtime{{\ifnum\hour<12 \global\amorpm={am}%
        \else\global\amorpm={pm}\advance\hour by-12 \fi
        \ifnum\hour=0 \hour=12 \fi
        \number\hour:\ifnum\minute<10 0\fi\number\minute\the\amorpm}}
\edef\militarytime{\number\hour:\ifnum\minute<10 0\fi\number\minute}

\def\draftlabel#1{{\@bsphack\if@filesw {\let\thepage\relax
   \xdef\@gtempa{\write\@auxout{\string
      \newlabel{#1}{{\@currentlabel}{\thepage}}}}}\@gtempa
   \if@nobreak \ifvmode\nobreak\fi\fi\fi\@esphack}
        \gdef\@eqnlabel{#1}}
\def\@eqnlabel{}
\def\@vacuum{}
\def\draftmarginnote#1{\marginpar{\raggedright\scriptsize\tt#1}}

\def\draftlabel#1{{\@bsphack\if@filesw {\let\thepage\relax
   \xdef\@gtempa{\write\@auxout{\string
      \newlabel{#1}{{\@currentlabel}{\thepage}}}}}\@gtempa
   \if@nobreak \ifvmode\nobreak\fi\fi\fi\@esphack}
        \gdef\@eqnlabel{#1}}
\def\@eqnlabel{}
\def\@vacuum{}
\def\draftmarginnote#1{\marginpar{\raggedright\scriptsize\tt#1}}

\def\draft{\oddsidemargin -.5truein
        \def\@oddfoot{\sl preliminary draft \hfil
        \rm\thepage\hfil\sl\today\quad\militarytime}
        \let\@evenfoot\@oddfoot \overfullrule 3pt
        \let\label=\draftlabel
        \let\marginnote=\draftmarginnote
   \def\@eqnnum{(\theequation)\rlap{\kern\marginparsep\tt\@eqnlabel}%
\global\let\@eqnlabel\@vacuum}  }

\def\numberbysection{\@addtoreset{equation}{section}
        \def\theequation{\thesection.\arabic{equation}}}

\def\underline#1{\relax\ifmmode\@@underline#1\else
        $\@@underline{\hbox{#1}}$\relax\fi}

\def\titlepage{\@restonecolfalse\if@twocolumn\@restonecoltrue\onecolumn
     \else \newpage \fi \thispagestyle{empty}\c@page\z@
        \def\thefootnote{\fnsymbol{footnote}} }

\def\endtitlepage{\if@restonecol\twocolumn \else  \fi
        \def\thefootnote{\arabic{footnote}}
        \setcounter{footnote}{0}}  %\c@footnote\z@ }
\relax

\numberbysection
\hybrid

\def\beq{\begin{equation}}
\def\eeq{\end{equation}}
\def\p{\partial}
\def\G{\Gamma}
\def\g{\gamma}
\def\s{\widetilde \theta}
\def\z{\zeta}

\def\a{\alpha}
\def\b{\beta}
\def\e{\varepsilon}

\def\R{{\cal R}}
\def\A{{\cal A}}
\def\B{{\cal B}}

\def\V{{\cal V}}
\def\D{{\cal D}}
\def\F{{\cal F}}

\def\M{{\cal M}}

\def\P{{\cal P}}
\def\X{{\cal X}}
\def\SP{{\cal S}}

\def\dim{{\rm dim}}
\def\res{{\rm res}}
\def\F{{\cal F}}

\def\e{\varepsilon}
\def \matrix #1 {\left(\begin{array}{cc} #1 \end{array}\right)}
\newtheorem{theo}{Theorem}[section]

\newtheorem{lem}{Lemma}[section]

\begin{document}
\begin{titlepage}
\title{Analytic theory of difference equations with rational and elliptic
coefficients and the Riemann-Hilbert problem}

\author{I.Krichever \thanks{Columbia University, New York, USA and
Landau Institute for Theoretical Physics, Moscow, Russia; e-mail:
krichev@math.columbia.edu. Research is supported in part by National Science
Foundation under the grant DMS-01-04621.}}

\date{July, 2004}

\maketitle

\begin{abstract} A new approach to the analytic theory of difference equations
with rational and elliptic coefficients is proposed. It is based on the construction
of canonical meromorphic solutions which are analytical along "thick paths".
The concept of such solutions leads to the notion of local monodromies of
difference equations. It is shown that in the continuous limit  they converge
to the monodromy matrices of differential equations. New type of isomonodromic deformations
of difference equations with elliptic coefficients changing the periods of elliptic
curves is constructed.
\end{abstract}
%\vfill

\end{titlepage}

\section{Introduction}

It is well-known that correlation functions of diverse statistical models,
gap probabilities in the Random Matrix Theory
can be expressed in terms of solutions of the Painlev\'e type {\it differential} equations
(see \cite{tr-wi,jmms,mal,hi,bd} and references therein).
In the recent years discrete analogs of the Painlev\'e equations
\cite{jsak,sak} have attracted considerable interest due to their connections to
discrete probabilistic models \cite{bor1,bb}.
In \cite{bor} it was found that the general setup for these equations is provided by
the theory of {\it isomonodromy} transformations of linear
systems of difference equations with rational coefficients.

The analytic theory of matrix linear difference equations
\beq\label{1}
\Psi(z+1)=A(z)\Psi(z)
\eeq
with rational coefficients is a subject of its own interest. It goes back to the
fundamental results of Birkhoff \cite{bi1,bi2} which have been developed later by many
authors (see the book \cite{put} and references therein).

Difference equations (\ref{1})
are classified in a rough way by terms: {\it regular, regular singular, mild and wild}
(see \cite{put} for details). The terminology reflects the formal asymptotic theory of
the equation near the infinity. Equation (\ref{1}) with the coefficients of the form
\beq\label{L}
A=A_0+\sum_{m=1}^n {A_m\over z-z_m}
\eeq
is regular singular if $A_0=1$. It is regular if in addition $A$ has no
residue at the infinity, i.e. $\sum_{m=1}^n A_m=0$. The mild equations are those for
which the matrix $A_0$ is invertible. In this paper we restrict ourself to the case of
mild equations with a diagonalizable leading coefficient $A_0$. It will
be assumed also that the poles $z_m$ are not congruent, i.e. $z_l-z_m$ is not an integer,
$z_l-z_m\notin Z$.

Equation (\ref{1}) is invariant under the transformation $\Psi'=\rho^{\,z}\Psi,\
A'=\rho A(z)$, where $\rho$ is a scalar. It is also
invariant under the gauge transformation $\Psi'=g\Psi,\
A'=gA(z)g^{-1}, \ g\in SL_r.$ Therefore, if $A_0$ is diagonalizable, then
we may assume without loss of generality that $A_0$ is a diagonal matrix
of determinant 1,
\beq\label{a0}
A_0^{ij}=\rho_i\delta^{ij},\ \ \det A_0=\prod_j \rho_j=1.
\eeq
In addition, throughout the paper it will be assumed that
\beq\label{trace}
{\rm Tr} \left({\rm res_{\infty}}Adz\right)={\rm Tr} \left(\,\sum_{m=1}^n A_m\right)=0.
\eeq
If the eigenvalues of $A_0$ are pairwise distinct $\rho_i\neq \rho_j$, then equation
(\ref{1})
has a unique formal solution $Y(z)$ of the form
\beq\label{Y1}
Y=\left(1+\sum_{s=1}^{\infty} \chi_s z^{-s}\right)e^{z\ln A_0+K\ln z},
\eeq
where $K^{ij}=k_i\delta^{ij}$ is a diagonal matrix.

In \cite{bi1,bi2} difference equations with {\it polynomial}
coefficients $\widetilde A$ were considered. Note, that the general case of
rational $A(z)$ is reduced to the polynomial one by the transformation
\beq\label{rp}
\widetilde A=A(z)\prod_m(z-z_m), \ \ \widetilde \Psi= \Psi\prod_m \G(z-z_m),\
\eeq
where $\G(z)$ is the Gamma-function. Birkhoff proved that, if the ratios
of eigenvalues $\rho_i$ of the leading coefficient of $\widetilde A$ are not real,
$\Im\, (\rho_i/\rho_j)\neq 0$, then equation (\ref{1}) with polynomial coefficients
has two canonical meromorphic solutions $\widetilde \Psi_r(z)$ and $\widetilde\Psi_l(z)$
which are holomorphic and asymptotically represented by $\widetilde Y(z)$ in
the half-planes ${\Re}\, z>>0$ and ${\Re}\, z<< 0$, respectively.
Moreover, Birkhoff proved that the connection matrix
\beq \label{S}
\widetilde S(z)=\widetilde\Psi_r^{-1}(z)\widetilde\Psi_l(z)\, ,
\eeq
which must be periodic for obvious reason, is, in fact, a rational function in
$\exp(2\pi iz)$. This function has just as many constants involved as there are
parameters in $\widetilde A$. The other result of Birkhoff implies that
if two polynomial matrix functions $\widetilde A'(z)$ and $\widetilde A(z)$
have the same connection matrix $S(z)$ then
there exists a rational matrix $R(z)$ such that
\beq\label{eq}
\widetilde A'(z)=R(z+1)\widetilde A(z)R^{-1}(z)\,.
\eeq
In \cite{bor} a family of commuting transformations (\ref{eq}) was explicitly
constructed. Furthermore, it was shown that in the continuous limit
the commutativity equations for a certain subset of these transformations
converge to the classical Schlesinger equations (\cite{schles}).

Until now key ideas of Birhoff's approach to the analytic theory of difference
equations have remained intact. A construction of actual solutions of (\ref{1})
having prescribed asymptotic behavior in various sectors at infinity
resembles rather the Stocks' theory of differential equations with
irregular singularities, then the conventional theory of differential
equations with regular singularities. The monodromy representation
of $\pi_1 (C\backslash \{z_1,\ldots,z_n\})$ which provides the integrals of motion for
the Schlesinger equations, has no obvious analog in
discrete situation. On the other hand the obvious differential analog
of the connection matrix $S(z)$ gives only the monodromy information at infinity
and provides no information on local monodromies around the poles $z_m$.
(Maybe by this reason Birkhoff just from the beginning eliminated the positions of poles
and restricted himself to the case of polynomial coefficients).

The main goal of this paper is to develop a new approach to the analytic theory
of difference equations with rational coefficients and extend it to the case
of equations with {\it elliptic} coefficients. It is based on the construction
of meromorphic solutions of difference equations which are holomorphic along
{\it thick paths}.

It is instructive to present the case just opposite to the Birkhoff's one,
namely, the case of real {\it exponents} $\rho_i$.
Let $x$ be a real number such that $x\neq \Re z_i$. Consider matrix solution
$\Psi_x(z)$ of equation (\ref{1}) that is non-degenerate and holomorphic inside the
strip $z\in \Pi_x: x\leq \Re z\leq x+1$ and continuous up to the boundary. It is
also required that in $\Pi_x$ the solution $\Psi_x$ grows at most {\it polynomially} as $|\Im z|\to \infty$.
It is easy to show that if such a solution exists then it is unique up to the
transformation $\Psi_x'=\Psi_x(z)g,\ \ g\in GL_r$. Moreover, it turns out that,
if $\Psi_x$ exists, then it has the following asymptotic representation
\beq
\Psi_x=Yg_x^{\pm},\ \ \Im z\to \pm \infty\ .
\eeq
To some extend, the ratio
\beq\label{g}
g_x=g_x^{+}\left(g_x^-\right)^{-1}
\eeq
can be regarded as a transfer matrix of the solution along the "thick"
path $\Pi_x$ from $-i\infty$ to $i\infty$.

Furthermore, we show that for $x>>0$ and $x<<0$ the solution $\Psi_x$ {\it does exist}.
In these regions it is $x$-independent. Therefore, we get two meromorphic solutions
$\Psi_r$ and $\Psi_l$ of equation (\ref{1}), which are holomorphic in the half-planes
$\Re z>>0$ and $\Re z<<0$, respectively. The corresponding transfer matrices $g_r=g_x,\,
x>>0$, and $g_l=g_x,\, x<<0$ are "quasi"-upper or -lower triangular matrices, i.e.
\beq\label{T}
g_{r(l)}^{ii}=1, \ g_{r}^{ij}=0,\ {\rm if}\ \rho_i<\rho_j,\ \
g_{l}^{ij}=0,\ {\rm if} \ \rho_i>\rho_j,
\eeq
This result clarifies the well-known fact that, if $\Im \, (\rho_i/\rho_j)=0$, then
there are no Birkhoff's solutions with uniform asymptotic representation in the half-planes
${\Re}\, z>>0$ and ${\Re}\, z<< 0$.

If the solutions $\Psi_r,\, \Psi_l$  are normalized by the condition
$g_x^-=1$, then their connection matrix has the form
\beq\label{S1}
S(z)=\Psi_r^{-1}(z)\Psi_l(z)=1-\sum_{m=1}^n {S_m\over e^{2\pi i(z-z_m)}-1},
\eeq
where
\beq\label{S0}
S_{\infty}=1+\sum_{m=1}^n\, S_m=g_r^{-1}e^{2\pi i K}g_l\,,
\eeq
and $K$ is the same diagonal matrix as in (\ref{Y1}). To the best of the author's knowledge,
the explicit form (\ref{S1}) of the connection matrix including the relations
(\ref{T},\ref{S0}) is a new result even for the case
of regular singular equations for which $g_{r(l)}=1$
(compare it with the Theorem 10.8 in \cite{put}).

The direct monodromy map
\beq\label{600}
A(z)\to S(z)
\eeq
for regular singular and mild equations is constructed in sections 2 and 3, respectively.
In section 2.2 we introduce a notion of {\it local monodromies} of difference equations.
First, they are defined for three examples of regular singular equations. Namely, for the
case of special equations with the coefficients $A\in \A_0$ of the form (\ref{L}) such that
$\det A(z)\equiv 1$. The second case considered is the case of unitary
difference equations with the coefficients
$A\in \A^U$ satisfying the relation $A^+(\bar z)=A^{-1}(z)$. The third example is the small
norm case, i.e. the case of equations with coefficients such that $|A_m|<\e$.

The existence of the canonical solution $\Psi_x$ is equivalent to  solvability
of an auxiliary system of linear singular equations. The index of that system
equals
\beq\label{ind}
ind_x A={1\over 2\pi i} \int_{L} d\ln \det A, \ \ z\in L: \Re z=x.
\eeq
Fundamental results of the theory of singular integral equations (\cite{mus})
imply that, if $ind_x A=0$ then for generic $A$ the canonical solution $\Psi_x$ exists.

The index $ind_x A$ vanishes identically, if $\det A=1$.
Therefore, for generic $A\in \A_0$ the solution $\Psi_x$ exists for all $x\neq \Re z_k$.
It is $x$-{\it independent}, when $x$ varies between the values $\Re \, z_k$.
Suppose that $\Re z_1<\Re z_2<\cdots<\Re z_n$, then we obtain a set of $n+1$ meromorphic
solutions $\Psi_k(z)$ of equation (\ref{1}) that are holomorphic in
the domains $\Re z_k<\Re z<\Re z_{k+1}+1$ (here $k=0,\ldots, n,$ and
for brevity we formally set $z_0=-\infty$ and $z_{n+1}=\infty$).

The local connection matrices $M_k=\Psi_{k}^{-1}\Psi_{k-1}$ have the form
\beq\label{sj}
M_k=1-{m_k\over e^{2\pi i(z-z_k)}-1}.
\eeq
The evaluation of $M_k$ at $z=i\infty$ equals
\beq\label{mu}
\mu_k=1+m_k=g_{k}^{-1}g_{k-1},
\eeq
where $g_k$ is the transfer  matrix (\ref{g}) along the strip $\Pi_x$ for
$\Re\, z_k<x<\Re\, z_{k+1}$. The matrix $\mu_k$ is a discrete analog
of the {\it monodromy matrix} along a path  from $-i\infty$ which goes around the puncture
$z_k$ and returns back to $-i\infty$.

The monodromy matrices $\mu_k$ uniquely define local
connection matrices $M_k(z)$ and the global connection matrix
(\ref{S1}), which is equal to the product \beq\label{sosj}
S(z)=M_n(z)M_{n-1}(z)\cdots M_1(z). \eeq Note, that a generic
unimodular matrix $S(z),\ \det S=1,$ of the form (\ref{S1}) has a
unique representation (\ref{sosj}), where the factors $M_k$ have
the form (\ref{sj}). Therefore, the correspondence
$S(z)\leftrightarrow \{\mu_k\}$ is one-to-one on open sets of the
corresponding spaces.

In all the three examples of difference equations considered in section 2.2,
we show that the monodromy map (\ref{600}) is one-to-one on open sets of the corresponding
spaces. The solution of the inverse monodromy problem of reconstruction of the coefficients
$A(z)$ from the monodromy data is reduced to a certain Riemann-Hilbert factorization problem
on a set of vertical lines. The possibility of this reduction is based on an existence
of intermediate solutions $\Psi_l,\Psi_1,\ldots,\Psi_r$,
whose domains of analyticity overlap and cover the whole complex plane.

For generic difference equations the inverse monodromy problem is solved in section 3.
We prove that the monodromy map restricted to the subspace $\A_D$
of coefficients having {\it fixed} determinant
\beq\label{delt0}
A\in \A_{D}\subset \A: \ \
\det A(z)=D(z)={\prod_{\a=1}^N (z-\z_{\a})\over \prod_{i=1}^n(z-z_m)^{h_m}},\ \
N=\sum_m h_m\, ,
\eeq
is a one-to-one correspondence of open dense sets.
If the zeros $\z_{\a}$ of $D$ are not congruent to each other, then
the injectivity of (\ref{600})  restricted to $\A_D$
follows directly from the construction of canonical solutions.
Isomonodromy transformations are used as an important
intermediate step for the proof of surjectivity of (\ref{600}).

Let us call two rational functions $D$ and $D'$
of the form (\ref{delt0}) equivalent
if their zeros and poles are pairwise congruent, i.e.
$\z_{\a}-\z'_{\a}\in Z, \ z_m-z_m'\in Z$. It turns out that for each pair of equivalent
functions there exists a birational isomonodromy isomorphism
$T_{D}^{D'}:\A_D\longmapsto {\A_D'}$. Therefore, in order to prove that
there is a map
\beq\label{invmap}
\SP_{\widehat D}\longmapsto \A_D,
\eeq
which is inverse to the restriction of (\ref{600}) to $\A_D$, it is enough
to construct (\ref{invmap}) for at least one $D$ in each equivalence class $[D]$.
Here $\SP_{\widehat D}$ is the space of connection matrices
having fixed determinant $\widehat D=D(w),\ \ w=e^{2\pi i z}$; and (\ref{invmap}) is
defined on an open dense subspace of $\SP_{\widehat D}$.

In each equivalence class $[D]$ there exists a representative $D$
such that its zeros and poles belong to $\Pi_x$. In that case the canonical
meromorphic solutions $\Psi_{l}$ and $\Psi_r$ are holomorphic in the domains
$\Re z<x+1$ and $\Re z>x$, which overlap. Then, the problem of reconstruction of
$\Psi_{r(l)}$ is reduced to the standard Riemann-Hilbert factorization problem
on the line $\Re z=x+1/2$.

In Section 4 we consider the continuous limit of our construction.
It turns out that the canonical meromorphic solutions $\Psi_x$
of a difference equation
\beq\label{1e}
\Psi(z+h)=\left(1+hA_0+h\sum_{m=1}^n {A_m\over z-z_m}\right)\Psi(z)
\eeq
exist for any $x$ such that $|x-\Re\, z_m|> Ch$.
Furthermore, we show that in the limit $h\to 0$
this solution in the neighborhood of the path $\Re\, z=x$ converges to
a solution of the differential system
\beq\label{1dif}
{d\widehat \Psi\over dz} =\left(\,A_0+\sum_{m=1}^n {A_m\over z-z_m}\right)\widehat\Psi(z).
\eeq
That implies that the monodromy matrices $\mu_k$ do converge to the conventional
monodromy matrices of the corresponding system of differential equations.
For difference equations with real exponents the transfer matrices $g_{r(l)}$
converge to the Stokes' matrices of equation (\ref{1dif}) at the infinity, where
the differential equation (\ref{1dif}) has irregular singularity. Similar result
is obtained for the Birkhoff's case of imaginary exponents.

In Section 5 we extend our consideration to the case of difference
equations with "elliptic" coefficients. More precisely, we consider the equation
\beq\label{el10}
\Psi(z+h)=A(z)\Psi(z),
\eeq
where $A(z)$ is a meromorphic $(r\times r)$ matrix function with simple poles,
which satisfies the following monodromy properties
\beq\label{elmon0}
A(z+2\omega_{\a})=B_{\a}A(z)B_{\a}^{-1},\ \ B_{\a}\in SL_r,\ \a=1,2.
\eeq
The matrix $A(z)$ can be seen as a meromorphic section of the vector bundle $Hom (\V,\V)$,
where $\V$ is a holomorphic vector bundle on the elliptic curve
$\G$ with periods $2\omega_{\a}$, which is defined by a pair of commuting
matrices $B_{\a}$. If $B_{\a}$ are diagonalizable then without loss of generality we may
assume that $\B_{\a}$ are diagonal. Furthermore, using the gauge transformations
defined by diagonal matrices of the form $G^z$ one can make $B_1$ to be equal to the identity
matrix. In this gauge the second matrix $B_2$ can be represented in the
form $B_2=e^{\pi i\hat q/\omega_1}$, where $\hat q$ is a diagonal matrix.

Without loss of generality we may assume that $\Im \left(\omega_2/h\right)>0$.
Along the lines identical to that in the rational case, we define the canonical
meromorphic solutions $\Psi_x$ of equation (\ref{el10}). They satisfy the following
{\it Bloch} monodromy property
\beq\label{bloch}
\Psi_x(z+2\omega_2)=e^{\pi i\hat q/\omega_1}\Psi_x(z)e^{-2\pi i\hat s/h},
\eeq
where $\hat s$ is a diagonal matrix $\hat s^{\,ij}=s_i\delta^{ij}$.
The connection matrix $S_x$ of two such solutions $\Psi_x(z)$ and
$\Psi_{x+1}(z)=\Psi_x(z-2\omega_1)$\,, i.e.
\beq\label{cone}
\Psi_x(z)=\Psi_{x}(z-2\omega_1)S_x(z)\,,
\eeq
has the following monodromy properties
\beq\label{mos}
S_x(z+h)=S_x(z),\ \ S_x(z+2\omega_2)=e^{2\pi i\hat s/h}S_x(z)e^{-2\pi i\hat s/h},
\eeq
and can be seen as a section of a bundle on the elliptic curve with periods $(h, 2\omega_2)$.

The correspondence $A(z)\to S_x(z)$ is a direct monodromy map in the elliptic case.
As in the rational case, a single-valued branches of the inverse monodromy map are
defined on subspaces of coefficients $A(z)$ with fixed determinant.
Isomonodromy transformations which change
the positions of poles and zeros of $A$ are constructed in a way similar to the rational
case. We also construct a new type of isomonodromy transformations which
{\it change the periods} of elliptic curves. These transformations have the form
\beq\label{ttt}
A'(z)=\R(z+h)A(z)\R^{-1}(z)\, ,
\eeq
where $\R$ is a meromorphic solution of the difference equation
\beq\label{er}
\R(z+2\omega_1+h)A(z)=\R(z),
\eeq
which has the following monodromy property
\beq\label{er1}
\R(z+2\omega_2)=e^{2\pi i\hat q\,'/(h+2\omega_1)}\R(z)e^{-\pi i\hat q/\omega_1}
\eeq
The existence of such transformations shows that in the elliptic case there is
a certain symmetry between the periods $2\omega_{\a}$ of an elliptic curve and the
step $h$ of the difference equation. Note, that this type of symmetry
for $q$-analog of the elliptic Bernard-Knizhnik-Zamolodchikov equations was
found in \cite{var}.

\section {Meromorphic solutions of difference equations and Riemann-Hilbert problem}

The matrix differential equation $\p_z \Psi=A(z)\Psi$ with rational coefficients
has multi-valued holomorphic solutions on $C\backslash \{z_m\}$, where $z_m$ are the poles of
$A(z)$. The initial condition $\Psi(z_0)=1,\ z_0\neq z_m,$ uniquely defines $\Psi$
in the neighborhood of $z_0$. This simple but fundamental fact is a starting point
of the analytical theory of differential equations with rational coefficients.
Analytic continuation of $\Psi$ along paths in $C\backslash \{z_m\}$ defines the monodromy
representation $\mu:\pi_1(C\backslash \{z_m\})\longmapsto GL_r .$

A construction of meromorphic solutions of the difference equations is less obvious.
It can be easily  reduced to a solution of the following auxiliary Riemann-Hilbert type problem.

\medskip
\noindent{\bf Problem I:} {\it To find in the strip $\Pi_x: \ x\leq \Re\, z\leq x+1,$ a
continuous matrix function $\Phi(z)$ which is meromorphic inside $\Pi_x$,
and such that its boundary values on the two sides of the strip satisfy the equation}
\beq\label{R1}
\Phi^+(\xi+1)=A(\xi)\Phi^-(\xi),\ \ \xi=x+iy\,.
\eeq

If $\Phi$ is a solution of this problem, then equation (\ref{1}) can be used
to extend it to a function $\Psi$ on the whole complex plane.
{\it A'priori} $\Psi$ is meromorphic outside the lines $\Re\, z=x+l, \ l\in Z$.
On these lines $\Psi$ is continuous due to (\ref{R1}). Recall a well-known property
of analytic functions: if $f$ is a continuous function in a domain $D$ of the plane
and is holomorphic in the complement $D\backslash L$ of a smooth arc $L$, then
$f$ is holomorphic in $D$. Therefore, $\Psi$ is meromorphic on the whole complex
plane and can be regarded as a meromorphic solution of (\ref{1}).

The function $t=\tan( \pi z)$ defines a one-to-one conformal map of the interior of $\Pi_x$
onto the complex plane of the variable $t$ with a cut between the punctures
$t=\pm 1$. Under this map the problem (\ref{R1}) gets transformed to the standard
Riemann-Hilbert factorization problem on the cut. Fundamental results of the theory of
singular integral equations imply that the problem (\ref{R1}) always has solutions. Moreover,
if the index (\ref{ind}) of the corresponding systems of singular integral equations
equals zero, then for a generic $A(z)$ this problem has {\it sectionally holomorphic}
non-degenerate solution. The later means that there exists a constant $\a<1$
such that $(t\pm 1)^{\a}\Phi(t)$ is bounded at the edges of the cut. In terms of the
variable $z$ a sectionally holomorphic solution $\Phi_x$ of the Problem I is a
non-degenerate holomorphic matrix function inside $\Pi_x$ and such that
\beq\label{gr}
\exists\ \  0 \leq\a<1, \ \ |\Phi(z)|< e^{2\pi \a |\Im\, z|},\ \ |\Im\,  z|\to \infty.
\eeq
This solution is unique up to the transformation $\Phi'(z)=\Phi(z)g, \  g\in SL_r$.

Almost all the results of this section do not require any additional information.
Let us provide some details needed for asymptotic description of $\Psi_x$.

\subsection{Regular singular equations}

We begin with the case of regular singular difference equation, i.e. equation
(\ref{1}) with the coefficient $A(z)$ of the form
\beq\label{rs}
A=1+\sum_{m=1}^n {A_m\over z-z_m}\,.
\eeq
Equation (\ref{1}) is invariant under the gauge transformations $A'=gAg^{-1}, \Psi'=g\Psi,\
g\in SL_r$. Thus, if the residue of $Adz$ at the infinity is diagonalizable,
we may assume without loss of generality that
\beq\label{res}
K=\res_{\infty}\, Adz=\sum_{m=1}^n A_m={\rm diag} (k_1,\ldots,k_r).
\eeq
If $k_i-k_j\notin Z$, then equation (\ref{1}) has a unique formal solution of the form
\beq\label{Y}
Y=\left(1+\sum_{s=1}^{\infty}\chi_sz^{-s}\right)z^K\, .
\eeq
The coefficients $\chi_s$ are defined by equations, which are obtained by
substitution of (\ref{Y}) into (\ref{1}). These equations express
$[K,\chi_{s}]+s\chi_{s}$ in terms of $A_i$ and $\chi_1,\ldots, \chi_{s-1},$ and can be
recurrently solved for $\chi_s$.

Let $\P_x$ be the space of continuous functions
$\Phi(z)$ in the strip $\Pi_x$, which are holomorphic inside
the strip and have at most polynomial growth at the infinity, i.e.
\beq\label{bound1}
\Phi\in \P_x: \ \exists\, N,\ \  \vert\, \Phi\, \vert<|\,z\,|^{N}, \ \ z\in \Pi_x.
\eeq
\begin{lem} Let $x$ be a real number such that $x\neq\Re z_j$. Then:

(a) for $|x|>>$0 there exists a unique up to normalization
non-degenerate solution $\Phi_x\in \P_x$ of the
Riemann-Hilbert problem (\ref{R1});

(b) for a generic $A(z)$ the solution $\Phi_x\in \P_x$ exists and is unique up
to normalization for all $x$ such that $ind_x A=0$;

(c) at the two infinities  of the strip the function $\Phi_x$ asymptotically equals
\beq\label{as}
\Phi_x(z)=Y(z)g_{x}^{\pm},  \ \ \Im z\to \pm \infty.
\eeq
\end{lem}
{\bf Remark.} Part (c) of the lemma means that, if
$Y_{m'}=\left(1+\sum_{s=1}^{m'}\chi_s z^{-s}\right)z^K$
is the partial sum of (\ref{Y}) then
\beq\label{as00}
\left|\, \Phi_x\left(Y_{m'} g_x^{\pm}\right)^{-1}-1 \right|\leq O(|z|^{-m'-1}),\
\Im\, z\to \pm \infty\,
\eeq
and the estimate (\ref{as00}) is uniform in the domain
$z\in \Pi_{x,\e}: \ x+\e\leq \Re\, z\leq x+1-\e$ for any $\e>0.$

{\it Proof.} First let us show that if $\Phi_x$ exists, then it is unique up to
the normalization.
The determinant of $\Phi_x$ is a holomorphic function inside $\Pi_x$.
Its boundary values on the two sides of the strip satisfy the relation:
$\ln \det \Phi_x^+(\xi+1)=\ln \det \Phi_x^-(\xi)+\ln \det A(\xi).$
If $ind_x A=0$, then the principal part of the integral of $(d\ln \det \Phi_x)$
along the boundary of $\Pi_x$ equals zero. Therefore, if $\Phi_x$ is non-degenerate at least
at one point then it is non-degenerate at all the points of $\Pi_x$.
Now, suppose that there are two solutions of the factorization problem,
then $g=\Phi_x^{-1}\Phi_x'$ is an entire periodic function.
It can be regarded as a function $g(w)$ of the variable $w=e^{2\pi iz}$,
holomorphic outside the points $w=0, \ w=\infty$.
From (\ref{bound1}) it follows that
\beq
\lim_{w\to 0} w g(w)=0, \ \lim_{w\to \infty} w^{-1} g(w)=0.
\eeq
Therefore, $g(w)$ has an extension which is holomorphic at
the points $w=0$ and $w=\infty$.
Thus it is a constant matrix.

In a standard way the problem (\ref{R1}) is reduced to the system of linear singular equations.
Let us fix for each positive integer $m$ a holomorphic in $\Pi_x$ function $Y_m$,
which at $\pm i\infty$ coincides with $Y$ up to the order $m$. If $0\notin \Pi_x$, then we
can define $Y_m$ by the $m$-th partial sum of (\ref{Y}). If $0\in \Pi_x$, then we choose
$x_0\notin \Pi_x$ and take $Y_m$ in the form
\beq\label{C}
Y_m=\left(1+\sum_{s=1}^m\widetilde \chi_s(z-x_0)^{-s}\right)(z-x_0)^{K}\, ,
\eeq
where the coefficients $\widetilde \chi_s$ are uniquely defined by the congruence
\beq\label{CY}
\left(1+\sum_{s=1}^m \widetilde \chi_s(z-x_0)^{-s}\right)\left({z-x_0\over z}\right)^{K}
\left(1+\sum_{s=1}^m\chi_s z^{-s}\right)^{-1}=1+O(z^{-m-1}).
\eeq
Each sectionally holomorphic in $\Pi_x$ function can be represented by
the Cauchy type integral. Let us consider a function $\Phi_x$ given by the formula
\beq\label{psi}
\Phi_x=Y_m\phi,\ \ \phi=1+\int_L
\varphi(\xi)k(z,\xi)\,d\xi,
\eeq
where $L$ is the line $\Re\, \xi=x$ and
\beq\label{ker}
k(z,\xi)=
{e^{\pi i(z-x)}+e^{-\pi i(z-x)}\over
(e^{\pi i (\xi-x)}+e^{-\pi i (\xi-x)})(e^{\pi i (\xi-z)}-e^{-\pi i (\xi-z)})}\ .
\eeq
Let $H$ be the space of H\"elder class functions on $L$, such that
\beq\label{hel}
\varphi\in H: \exists\,  \a<1,\ |\,\varphi(\xi)|<O(e^{\pi \a |\Im\, \xi|}).
\eeq
If $\varphi\in H$, then the integral in (\ref{psi}) converges and defines a function $\phi$, which is
holomorphic inside $\Pi_x$, and is continuous up to the boundary.
The boundary values $\phi^{\pm}$ of $\phi$ are
given by the Sokhotski-Plemelj
formulae
\beq\label{psi1}
\phi^-(\xi)=
1+I_{\varphi}(\xi)-{\varphi(\xi)\over 2},\ \
\phi^+(\xi+1)=1+I_{\varphi}(\xi) +{\varphi(\xi)\over 2},
\eeq
where $I_{\varphi}(\xi)$ denotes the principle value of the integral
\beq\label{phi}
I_{\varphi}(\xi)=p.v.\int_L
\varphi(\xi')k(\xi,\xi')\,d\xi'\,.
\eeq
The equation (\ref{R1}) is equivalent to the following
nonhomogeneous singular integral equation
\beq\label{K}
(\widetilde A+1)\,\varphi-2(\widetilde A-1)I_{\varphi}=2(\widetilde A-1).
\eeq
where
\beq\label{am}
\widetilde A=Y_m(\xi+1)^{-1}A(\xi)Y_m(\xi).
\eeq
By definition of $Y_m$, for large $|\,z|$
we have
\beq\label{am1}
|\widetilde A(\xi)-1|\leq O\left(|\xi|^{-m+\kappa}\right),\ \ \kappa={\rm max}_{ij}|k_i-k_j|.
\eeq
For large $|x|$ the left hand side of (\ref{am1}) is uniformly bounded by
$O\left(|x|^{-m+\kappa}\right)$, and equation (\ref{K}) can be solved by iterations.

Consider a sequence of the functions $\varphi_n$ defined recurrently
by the equation
\beq\label{Ka}
(\widetilde A+1)\,\varphi_{n}-2(\widetilde A-1)I_{\varphi_{n-1}}=2(\widetilde A-1),
\eeq
where $\varphi_0=0$. For $n>0$ equation (\ref{Ka}) implies
\beq\label{Kb}
(\widetilde A+1)\,(\varphi_{n+1}-\varphi_n)=2(\widetilde A-1)I_{(\varphi_n-\varphi_{n-1})}.
\eeq
Therefore, if the norm of $(\widetilde A-1)$ is small enough, then
$|\varphi_{n+1}-\varphi_n|<c\,\e^n,\ \e<1.$ The sequence $\varphi_n$  obviously converges to
a continuous function $\varphi$, which is a solution of (\ref{K}). Moreover, by standard
arguments used in the theory of boundary value problems (see \cite{mus} for details) it can be
shown that $\varphi$ is a H\"elder class function, and thus
the first statement of the Lemma is proven.

For any $x$ the left hand side of (\ref{K}) is a singular integral operator
$K: H\longmapsto H$. It has the Fredholm regularization. Furthermore,
the fundamental results of the theory of the Fredholm equations imply that
nonhomogeneous linear equation (\ref{K}) is solvable
if the adjoint homogeneous equation
\beq\label{Kd}
f(\xi)(\widetilde A(\xi)+1)-2\left(p.v.\int_L
f(\xi')k(\xi',\xi)d\xi'\right)(\widetilde A(\xi)-1)=0,
\eeq
for a (row) vector-function $f\in H_0$ has no solutions (see \S 53 \cite{mus}).
Here $H_0$ is the space of the H\"elder class functions that are {\it integrable} on $L$.
Each solution of (\ref{Kd}) defines the (row) vector-function
\beq\label{psid}
F(z)=\cos^2(\pi(z-x))\left(\int_L
f(\xi)k(\xi,z)\, d\xi\right)Y_m^{-1}(z),
\eeq
which is a solution of the dual factorization problem in $\Pi_x$
\beq\label{R2}
F(\xi+1)A(\xi)=F(\xi),\ \ \xi\in L .
\eeq
The Cauchy kernel $k(\xi,z)$ has a simple pole at $x'=x+1/2$. Therefore, $F$
is holomorphic inside $\Pi_x$ and equals zero at $x',\ F(x')=0$.
It is bounded as $|\Im z|\to \infty.$
Non-existence of such solution $F$ is an open condition.
That implies the second statement of the lemma.

From (\ref{K}, \ref{am1}) it follows that $I_{\varphi}$ is bounded at the infinity and
$|\varphi(\xi)|<O(|\xi|^{-m+\kappa})$. Let us show that for $z\in \Pi_{x,\e}$
\beq\label{am2}
\phi (z)=g^{\pm}+O(|z|^{-m+\kappa+1})\, , \  \Im\, z\to\pm \infty,
\eeq
where
\beq\label{am3}
g^{\pm}=1-{1\over 2}\int_L(\tan(\pi i(\xi-x))\pm 1)\, \varphi(\xi)\, d\xi.
\eeq
Consider the case $\Im z\to \infty$. The integral in (\ref{psi}) can be represented
as the sum of two integrals $I_1$ and $I_2$. The first one is taken over the interval
$L_1:(x-i\infty, \xi_0)$ and the second one over the interval $L_2:(\xi_0,x+i\infty)$,
where $\xi_0=x+i\Im z/2$. In $\Pi_{x,\e}$ the Cauchy kernel is uniformly bounded $k(z,\xi)<C$.
Therefore,
\beq\label{int1}
|I_2|<C\int_{L_2}|\varphi(\xi)|\, d\xi < O(|z|^{-m+\kappa+1}).
\eeq
For $\xi\in L_1$ we have $|\xi-z|>\Im z/2$. Therefore,
\beq\label{ker1}
k(z,\xi)=k_+(\xi)(1+O(e^{-\pi |z|})),\
k_+(\xi)=\left(1-\tan(\pi i(\xi-x)\right),\ \xi\in L_1
\eeq
Hence,
\beq\label{int2}
\left|I_2+1-g^+\right|<\left|\int_{L_2}k_+(\xi)\, \varphi(\xi)d\xi\right|
+O(e^{-\pi |z|})\int_{L}k_+(\xi)\, |\varphi(\xi)|d\xi<O(|z|^{-m+\kappa+1})
\eeq
The proof of (\ref{am2}) for $\Im z\to -\infty$ is identical.

The solution of the factorization problem is unique. Therefore, the left hand side
of (\ref{psi}) does not depend on $m$. Equation (\ref{am2}) implies (\ref{as00})
for $m'<m-2\kappa$. Now letting $m\to \infty$ we obtain
that (\ref{as00}) is valid for any $m'$ and the proof of the lemma is completed.

\begin{theo} If $A_0=1$ and $k_i-k_j\notin Z$, then:

(A) there are unique meromorphic solutions $\Psi_l$ and $\Psi_r$ of equation (\ref{1})
which are non-degenerate, holomorphic, and asymptotically represented by $Y(z)$
in the domains $\Re z<<0$ and $\Re z>>0$, respectively
\footnote{In the asymptotic equalities $\Psi_{r(l)}=Y$
we assume the choice of the single-valued branch of $\ln z$ on $C$
with a cut $arg \,z=\pi/2$};

(B) the  matrix $S=\Psi_{l}^{-1}\Psi_r$ has the form
\beq\label{S01}
S(z)=1-\sum_{m=1}^n {S_m\over e^{2\pi i(z-z_m)}-1},\ \
S_{\infty}=1+\sum_{m=1}^n\, S_m=e^{2\pi i K}\,,
\eeq
\end{theo}
The first statement and the form of the connection matrix $S(z)$
are known (see Theorem 10.8 in \cite{put}). The author has not found in literature
an explicit form of the matrix $S_{\infty}$. Birkhoff proved that $S_{\infty}=1$
for the regular equations, where $K=0$. In \cite{put} it is stated only that $S_{\infty}$ is
non-degenerate.

\medskip
\noindent
{\it Proof}. The function $\Phi_x$, when it exists, defines a meromorphic
solution $\Psi_x$ of the difference equation (\ref{1}).
\begin{lem} Let $x<y$ be real numbers such that the corresponding boundary problems
(\ref{R1}) have solutions in $\P_x$ and $\P_y$, respectively. Then function
$M_{x,y}=\Psi_y^{-1}\Psi_x$ has the form
\beq\label{monxy}
M_{x,y}=1-\sum_{k\in \,J_{x,y}} {m_{k,(x,y)}\over e^{2\pi i (z-z_k)}-1},
\eeq
where the sum is taking over a subset of indices $J_{x,y}$ corresponding to the poles
such that $x<\Re z_k< y$.
\end{lem}
{\it Proof.} By definition $\Psi_x$ is holomorphic in $\Pi_x$. In the domain $\Re z> x+1$
it has poles at the points $z_k+l, \ l=1,2\ldots,$ for $\Re\, z_k>x$. Therefore,
the function $M_{x,y}$ in $\Pi_y$ has poles at the points congruent to
$z_k, \ k\in\, J_{xy}$. The function $M_{x,y}$ is a periodic function of $z$.
The same arguments, as ones used above for the proof of the uniqueness of $\Phi_x$,
show that $M_{x,y}(w)$ considered as a function of the variable
$w=e^{2\pi iz}$ has holomorphic extension to the points
$w=0, \ w=\infty$. Hence $M_{x,y}(w)$ is a rational function of the variable $w$.
It equals $1$ at $w=0$ and has poles at the points $w_k=e^{2\pi iz_k}, \
k\in\, J_{xy}$. Therefore, $M_{x,y}$ has the form (\ref{monxy}).

\medskip
\noindent
{\bf Remark.} The proof of the Lemma shows also that the existence of $\Phi_x$ for
a generic $A$ and $x$ such that $ind_x A=0$ is a simple direct corollary
of the existence of $\Psi_l$. Indeed, let $M_x$ be a function of the form (\ref{monxy}),
where the sum is taken
over all $z_k:\ \Re z_k<x$. Then the condition that the function $\Psi_x=\Psi_lM_x^{-1}$,
is holomorphic in $\Pi_x$ is equivalent to a system of algebraic equations on
the residues of $M_x$. If $ind_x A=0$, then the number of equations is equal to the
number of unknowns. Therefore, for a generic $A$ the canonical meromorphic
solution $\Psi_x$ of (\ref{1}) does exist.

The Lemma implies that $\Psi_x$ is locally $x$-independent.
In particular, $\Psi_x$ is $x$-independent in the infinite interval
$x< {\rm min}_k\{\Re z_k\}$. The corresponding function $\Psi_l$ is the unique
meromorphic solution of equation (\ref{1}), which is holomorphic at $\Re\, z<<0$
and asymptotically represented by $Y$, when $\Im z\to -\infty$, and asymptotically
represented by $Yg_l$, when $\Im z\to \infty$. For large $|x|$ the coefficient
$(\widetilde A-1)$ of equation (\ref{K}) is uniformly bounded. Therefore,
$\varphi$, which decays as $|\varphi(\xi)|<O(|\xi|^{-m+k})$ at the two edges of $L$
is also uniformly bounded by $O(|x|^{-m+k})$. Then from equation (\ref{am3})
it follows that $g_x^{\pm}=1+O(|x|^{-m+k})$. The matrix $g_l=g_x^+\left(g_x^-\right)^{-1}$
is $x$-independent. Hence, $g_l=1$ and $\Psi_l$ is asymptotically represented
by $Y$ in the whole half-plane $\Re z<< 0$. The same arguments show that $\Psi_x$
for $x>>0$ can be identified with $\Psi_r$. The statement $(A)$ of the theorem is proved.

The formula (\ref{S01})  is a particular case of the formula (\ref{monxy}).
In order to complete the proof of the statement $(B)$, we
recall that the definition of $Y$, and therefore, the normalization of $\Psi_x$
requires to fix a branch of $\ln z$. In our consideration it was always fixed
on the $z$ plane with a cut along the positive half of the imaginary axis.
In this case, the evaluation of $S$ at $-i\infty$ equals $1$, and its evaluation
at $i\infty$ is equal to the ratio of $z^K$ on two edges of the cut.

\subsection{Local monodromies.}

The necessary condition for the existence of a solution $\Phi_x$ of the boundary value problem
(\ref{R1}) is the equation $ind_x A=0$. If this condition is satisfied for all the values
of $x$ then we define a notion of local monodromies of difference equations (\ref{1}).

\bigskip
\noindent{\bf Special regular singular equations.}
We call regular singular equation (\ref{1})
{\it special}, if the residues $A_i$ of $A(z)$ are rank 1 matrices
\beq\label{aj}
A(z)=1+\sum_{k=1}^n {p_kq_k^T\over z-z_k}
\eeq
and the determinant of $A$ identically equals $1$, $\det\, A(z)=1.$
Here $p_k,q_k$ are $r$-dimensional vectors, considered modulo transformations
\beq\label{gauge}
p_k\to c_kp_k,\ \ \ q_k\to c_k^{-1}q_k\, ,
\eeq
where $c_k$ are scalars. The space of such matrices is of dimension $2N(r-2)$ and
will be denoted by $\A_0$. Explicit parameterization of an open set of
the space $\A_0$  can be obtained, if we order the poles, and represent $A(z)$
in the multiplicative form
\beq\label{A1}
A(z)\in \A_0:\ A(z)=\left(1+{a_nb_n^T\over z-z_n}\right)\cdots
\left(1+{a_1b_1^T\over z-z_1}\right)\, ,
\eeq
where $a_k,b_k$ are pairs of orthogonal vectors
\beq\label{deta}
b_k^Ta_k=0\, ,
\eeq
considered modulo the transformation (\ref{gauge}).
Equation (\ref{deta}) implies
\beq\label{det2}
\left(1+{a_kb_k^T\over z-z_k}\right)^{-1}=\left(1-{a_kb_k^T\over z-z_k}\right)\longmapsto
\det \left(1+{a_kb_k^T\over z-z_k}\right)=1.
\eeq
From (\ref{A1}, \ref{deta}) it follows that the parameters $p_k,q_k$ in
the additive representation (\ref{aj}) of $A$ satisfy the constraints
\beq\label{ver}
q_k^Tl_k^{-1}p_k=0,\ \ \ l_k=1+\sum_{m\neq k}^N {p_mq_m^T\over z_k-z_m}\, .
\eeq
For matrices $A\in \A_0$ the gauge fixing assumption (\ref{res})
has the form
\beq\label{resa}
\sum_{m=1}^n p_mq_m^T=\sum_{m=1}^n a_mb_m^T={\rm diag} \, (k_1,\ldots, k_r)=K\, .
\eeq
It is assumed throughout this subsection that the real parts of
the poles $r_k=\Re\, z_k$ are distinct, and $r_k<r_m,\ k<m$. For further use,
we introduce also the notation $r_0=-\infty, r_{n+1}=\infty$.
\begin{theo} $(i)$ For a generic matrix $A\in \A_0$, satisfying (\ref{resa}),
where $k_i-k_j\notin Z$, the corresponding special regular singular equation (\ref{1})
has a set of unique meromorphic solutions $\Psi_k,\ k=0,\ldots,n$, which are holomorphic
in the strips $r_{k}<\Re\, z<r_{k+1}+1$ and
asymptotically represented by $Yg_k^{\pm}$,\ as $\Im\, z\to \pm\infty$,
where $g_k^-=1$.

$(ii)$ The local connection matrices $M_k=\Psi_k^{-1}\Psi_{k-1}, \ k=1,\ldots, n,$ have the form
\beq\label{m25}
M_k=1-{\a_k\b_k^T\over e^{2\pi i (z-z_k)}-1}\ \
\eeq
where $(\a_k,\b_k)$ are pairs of orthogonal vectors
\beq\label{detb}
\b_k^T\a_k=0,
\eeq
considered modulo transformations (\ref{gauge}) and such that
\beq\label{resb}
(1+\a_n\b_n^T)\cdots (1+\a_1\b_1^T)=e^{2\pi i K}
\eeq

$(iii)$ The map of pairs of orthogonal vectors $\{a_m,b_m\}\longmapsto \{\a_k,\b_k\}$,
considered modulo transformation (\ref{gauge}), is a one-to-one correspondence of open sets
of the varieties defined by the constraints (\ref{deta}, \ref{resa}) and (\ref{detb},
\ref{resb}), respectively.
\end{theo}
{\it Proof.} As it was shown above, a solution $\Phi_x\in \P_x$
of the factorization problem (\ref{R1}) exists if the homogeneous singular integral
equation (\ref{Kd}) has no solutions. That is an open type condition and
therefore for generic $A$ the corresponding meromorphic solution $\Psi_x$ of
equation (\ref{1}) does exist. If $r_{k}<x<r_{k+1}$, then equation (\ref{1}) implies that
$\Psi_x$ has poles at the points $z_m+l, \ m=1,2\ldots,$ for $k<m$ and at the
points $z_m-l,\ l=0,1,\ldots,$ for $m\leq k$. Therefore, $\Psi_x$ is holomorphic in the strip
$r_{k}<\Re\, z< r_{k+1}+1$ and can be identified with $\Psi_k$. The solutions $\Psi_k$
exists for all $k$, if $A$ belongs to the intersection of open sets corresponding
to each $k$. It is still an open condition, therefore, $\Psi_k$ do exist for a generic
$A$. They are unique and have asymptotic representation, described in $(i)$.

The residues of $A(z)$ are rank 1 matrices. Therefore, the residue of $M_k$ at
$z_k$ is also a rank 1 matrix and can be represented in the form $\a_k\b_k^T$,
where $\a_k,\b_k$ are vectors defined up to the transformation (\ref{gauge}).
Then (\ref{monxy}) implies equation (\ref{m25}).
From the constraint $\det A=1$ and the normalization $g_k^-=1$ it follows that
$\det \Psi_k=1$. Hence, $\det M_k=1$. That implies (\ref{detb}).
The global connection matrix is the product of local ones,
$S=M_n\cdots M_1$. Therefore, equation (\ref{S01}) implies (\ref{resb}) and thus
the second statement of the Theorem is proven.

Now let us show that the map $\{a_m,b_m\}\longmapsto \{\a_k,\b_k\}$ is injective on
the open set of matrices $A\in \A_0$ for which the corresponding difference
equation has a set of canonical solutions $\Psi_k$.
Indeed, suppose that there exist two special regular singular equations
having the same local connection matrices. Then we have two sets of the corresponding
meromorphic solutions
$\Psi_k$ and $\Psi_k'$ which are holomorphic in the strips
$\Re\, z\in (r_k,r_{k+1}+1)$, and which are
asymptotically equal to $O(1)z^Kg_k^{\pm}, \ g_k^-=1$ as $\Im z \to \pm \ \infty$.
Note, that the matrices $g_k^{+}$ are the same for $\Psi_k$ and $\Psi_k'$ because
they are equal to the products of {\it the monodromy matrices} $\mu_k=1+\a_k\b_k^T$
\beq\label{m51}
g_0^+=1,\ \ g_k^+=\mu_{k-1}\cdots,\mu_1,\ k>1\, .
\eeq
The matrix function, which equals $\Psi_k'\Psi_k^{-1}$ in each of the corresponding
strips is continuous across the boundaries. Hence, it is an entire
function which is bounded at the infinity. It tends to $1$ as $\Im z\to -\infty$.
Therefore, it equals $1$ identically.

The proof of a surjectivity of the map $\{a_m,b_m\}\longmapsto \{\a_k,\b_k\}$
on an open set of the connection matrices once again is reduced to the Riemann-Hilber type factorization problem.
Let us fix a small enough real number $\varepsilon$. Then, the vertical lines $L_m:
\Re\, \xi=\Re\, z_m+\varepsilon$ divide the complex plane into $(n+1)$ domains
$\D_k, \ k=0, \ldots, n.$

\medskip
\noindent{\bf Problem II:} {\it For a given set of matrix functions $\M_j(\xi)$ on
$L_j$ find matrix functions $\X_k(z)$, which are holomorphic inside
the domains $\D_k$, continuous up to the boundaries, and whose boundary
values satisfy the equation }
\beq\label{RII}
\X_{k-1}^+(\xi)=\X_k^-(\xi)\M_k(\xi),\ \ \xi\in L_k\,.
\eeq

\bigskip
Let $M_k$ be a set of matrices of the form (\ref{m25}) satisfying the constrains
(\ref{detb},\ref{resb}). Then we consider first the Problem II for the set
of piece-wise constant matrices
\beq\label{m52}
\M_k^0(\xi)=1,\ \Im\, \xi\geq 0,
\ \M_k^0(\xi)=\mu_k,\  \Im\, \xi<0.
\eeq
This is just the inverse monodromy problem for differential equation, solved by Plemelj.
He showed that the solution of this problem exists if at least one of the monodromy matrices
is diagonalizable \cite{plem}. Let $\F_k$ be a solution of this auxiliary problem.
Then we define a new set of functions $\M_k(\xi)$ by the formula
\beq
\M_k=\F_k^+M_k\left(\F_k^-\right)^{-1}.
\eeq
The function $M_k$ tends to $\mu_k$ exponentially, as $\Im z\to \infty$. Therefore,
$\M_k\to 1$ at both the edges
of $L_k$. In that case we may find a solution of the problem (\ref{RII})
in the form of the Cauchy integral
\beq\label{bX}
\X(z)=1+\sum_k\int_{L_k} {\chi_k(\xi)d\xi\over \xi-z}\, .
\eeq
Inside each of the domains $\D_k$ formula (\ref{bX}) defines a holomorphic function
$\X_k$. Using the Sokhotski-Plemelj formulae for their boundary values we obtain
the system of singular integral equations for $\chi_k$
\beq\label{Kx}
{1\over 2}\chi_k(\xi)(\M_k(\xi)+1)-{1\over 2\pi i}I_{\chi}(\xi)(\M_k(\xi)-1)=(\M_k(\xi)-1),
\eeq
where $I_{\chi}(\xi)$ denotes the principle value of the integral
\beq\label{phi1}
I_{\chi}(\xi)=p.v.\sum_k\int_{L_k}
{\chi_k(\xi')d\xi'\over \xi'-\xi}\, .
\eeq
The non-homogeneous term of the system tends to zero at the infinity. Therefore, for a
generic set of matrices $M_k$ the system has a solution in the space of H\"elder class functions
decaying at infinity. That implies that $\X_k$ tends to the identity matrix at the infinity.
The functions $\F_k$ have asymptotic behavior $O(1)z^Kg_k^{\pm}$. Hence the functions
$\Psi_k=\X_k\F_k$ have the same asymptotic behavior. Its boundary values
satisfy the relation
\beq\label{RIIa}
\Psi_{k-1}^+(\xi)=\Psi_k^-(\xi)M_k(\xi),\ \ \xi\in L_k\,.
\eeq
This equation can be used for the meromorphic extension of $\Psi_k$ on the whole complex plane.
At the same time it shows that the function $A_k(z)=\Psi_k(z+1)\Psi_k^{-1}(z)$
is $k$-independent. In the domain $\D_k$ it has
a unique simple pole at $z_k$. Therefore, $A(z)$ is a meromorphic function with simple
rank 1 poles at the points $z_k$. It tends to the identity matrix at the infinity and $\det A=1$, i.e.
$A\in \A_0$ and thus the Theorem is proven.

\medskip
\noindent
{\bf Unitary difference equations.} As it has been emphasized above, for a given real number
 $x$ the canonical meromorphic solution $\Psi_x$ exists only for generic difference equations.
Here is an example of the class of difference equations for which the canonical solutions
{\it always exist}.

We call the difference equation unitary, if
its coefficient satisfies the condition
\beq\label{unit}
A(z)\in \A^U:\ \ A^+(\bar z)=A^{-1}(z)\, ,
\eeq
where $A^+$ is the hermitian conjugate of $A$.
An open set of such matrices can be parameterized by the sets of unit vectors $a_k$
\beq\label{unitpar}\
A(z)=\prod_{k=1}^n\left(1+a_ka^+_k{z_k-\bar z_k\over z-z_k}\right), \ \ a_k^+a_k=|a_k|^2=1.
\eeq
The factors in the product (\ref{unitpar}) are ordered so that the indices increase
from right to left. Recall that in this section we assume that the residue of $A$ at the
infinity is a diagonal matrix
\beq\label{resc}
\sum_{k=1}^n (z_k-\bar z_k)a_ka_k^+=K,\ \ K^{ij}=k_i\delta ^{ij},\ \ k_i-k_j\notin Z.
\eeq
Equation (\ref{unit}) implies ${\det \bar{A}(\bar z)}=\det A^{-1}(z)$.
Therefore, for any $x\neq \Re \, z_k$ the index of the boundary problem (\ref{R1})
equals zero, $ind_x\, A=0$.

\begin{lem} Let $A(z)$ be the coefficient of a regular singular unitary equation.
Then for each $x\neq \Re\, z_k$ the boundary problem (\ref{R1}) has
non-degenerate solution $\widetilde\Phi_x\in \P_x$ such that
\beq\label{u1}
\widetilde \Phi_x^+(\bar z)=\widetilde \Phi_x^{-1}(z).
\eeq
This solution is unique up to a unitary normalization
\beq\label{u2}
\widetilde\Phi_x'(z)=\widetilde\Phi_x(z)\, u,\ \ u\in U(r)\, .
\eeq
\end{lem}
{\it Proof.} As it was shown above, the Riemann-Hilbert problem
(\ref{R1}) has a solution $\Phi\in P_x$, if the dual boundary problem (\ref{R2}) has
no vector solution which is bounded at $-i\infty$ and
tends to zero  faster then any negative power of $\Im\, z$ at the other edge of the strip.
Suppose that such vector solution $F$ exists. Then the scalar function
$F(z)F^+(\bar z)$ is holomorphic in $\Pi_x$ and tends to zero at both edges of the
strip. Therefore, the integral of this function over the boundary
of the upper half $\Pi_x^+$ of the strip $\Pi_x$ exists and equals zero,
\beq\label{iherm}
\oint_{\p \Pi_x^+} F(z)F^+(\bar z)dz=0,\ \ \ z\in \Pi_x^+\subset \Pi_x: \, \Im\, z\geq 0.
\eeq
On the other hand, from (\ref{unit}) it follows that this function is periodic, i.e.
its evaluations at $\xi=x+iy$ and $\xi+1$ are equal.
Therefore, the integral (\ref{iherm}) equals the integral over the bottom
edge of $\Pi_x^+$
\beq\label{iherm1}
\oint_{\p \Pi_x^+} F(z)F^+(\bar z)dz=\int_x^{x+1}|F(x')|^2dx'>0.
\eeq
The contradiction of (\ref{iherm}) and (\ref{iherm1}) implies that $\Phi_x$ exists.
It was shown earlier that $\Phi_x$ is unique up to normalization. Let us normalize it by
the condition that $\Phi_x$ has asymptotic $Y$
as $\Im\, z\to -\infty$. At the other edge of the strip
it has asymptotic $Yg_x$ (in this subsection we don't use notations $g_x^{\pm}$ in order
to avoid confusing them with the sign of the hermitian conjugation.)

Our next goal is to show that $g_x$ is a positively defined hermitian matrix.
Indeed, from (\ref{unit}) it follows that if $\Phi_x$ is a solution of the boundary
problem, then the matrix $\left(\Phi_x^+(\bar z)\right)^{-1}$ is also a solution
of the same problem. That implies
\beq\label{u3}
\left(\Phi_x^+(\bar z)\right)^{-1}=\Phi_x(z)h, \ \ h\in GL_r.
\eeq
The evaluation of this equality at two edges of the strip gives $gh=1$ and $g^+h=1$.
Hence, $g=g^+$. The matrix $\Phi_x^+(\bar z) \Phi_x(z)$
is holomorphic in $\Pi_x$ and has equal values on two sides of the strip.
Hence, for any vector $v$ we have
\beq\label{u4}
\oint_{\p \Pi_x^+} v^+\, \Phi_x^+(\bar z)\Phi_x(z)\, vdz=0
\ \longmapsto \ v^+\,g\, v=
\int_x^{x+1}v^+\, \Phi_x^+(x')\Phi_x(x')\, vdx'>0.
\eeq
Thus $g$ is positively defined, and therefore there exists a matrix $g_1$ such
that $g=g_1^+g_1$. Equation (\ref{u3}) implies that the function
$\widetilde \Phi_x=\Phi_x g_1^{-1}$ satisfies (\ref{u1}).
\begin{theo} Let $A(z)$ be a matrix of the form (\ref{unitpar}). Then:

$(i)$ the corresponding
difference equation (\ref{1}) has a unique set of meromorphic solutions $\widetilde \Psi_k$,
such that: $(a)$ $\widetilde \Psi_k$ is holomorphic in the strip $r_{k}<\Re\, z<r_{k+1}+1$,
and grows at most polynomially, as $\Im\, z\to \pm\infty$; $(b)$
$\widetilde \Psi_0=\left(1+O(z^{-1})\right)z^K, \ \Im\, z\to -\infty$;
$(c)$ $\widetilde \Psi_k$ satisfies the relation
\beq\label{g+}
\widetilde \Psi_k^+(\bar z)=\widetilde \Psi_k^{-1}(z)\, ;
\eeq
$(d)$ the local connection matrices $M_k=\widetilde\Psi_k^{-1}\widetilde\Psi_{k-1}$
have the form
\beq\label{m25u}
\widetilde M_k(z)=1-f_k(z)\a_k\a_k^+,
\eeq
where
\beq\label{f}
f_k(z)=\left(1+|w_k|\right){ww_k^{-1}-|w_k|^{-1}\over ww_k^{-1}-1
},\ w=e^{2\pi iz},\ w_k=w(z_k)\,
\eeq
and $\a_k$ are unit vectors, $\a_k^+\a_k=1,$ satisfying the constraint
\beq\label{resu}
(1-\nu_n\a_n\a_n^+)\cdots (1-\nu_1\a_1\a_1^+)=e^{\pi i K},\ \ \nu_k=1+|w_k|.
\eeq
$(ii)$
The monodromy map of sets of unit vectors $\{a_k\}\longmapsto \{\a_k\}$
is a one-to-one correspondence of the varieties defined by equations (\ref{resc}) and
(\ref{resu}).

\end{theo}
{\it Proof.} Lemma 2.3 implies that solutions $\widetilde\Psi_k'$
satisfying conditions $(a)$ and $(c)$ exist and are unique up to
normalization. The corresponding connection matrix $\widetilde M_k'$, which is a
rational function of $w$, satisfies
the equation
\beq\label{uM}
\widetilde M_k'^+(\bar z)=\widetilde M_k'^{-1}(z)
\eeq
and has the only pole at $w_k$, where its residue is a rank $1$ matrix.
It is easy to check that each matrix, which satisfies these properties
has a unique representation in the form $\widetilde M_k'=u_k \widetilde M_k$,
where $\widetilde M_k$ is given by (\ref{m25u}) and $u_k\in U(r)$.
The condition $(b)$ uniquely normalizes $\widetilde \Psi_0$. Then, under
the change of the normalization $\widetilde\Psi_k'=\widetilde\Psi_ku_k,\ \ u_k\in U(r),$
the local connection matrices get transformed to $\widetilde M_k$.

The global connection matrix $\widetilde S=\widetilde M_n\cdots\widetilde M_1$
up to a $z$-independent factor is equal to the global
connection matrix $S$ corresponding to the canonically normalized solutions $\Psi_k$
used before, i.e. $\widetilde S=\widetilde S(-i\infty)S(z)$.
Therefore, using (\ref{uM}) we get
$S(i\infty)=\widetilde S^{-1}(-i\infty)\widetilde S(i\infty)=\widetilde S^2(i\infty)$.
The left hand side of (\ref{resu}) equals $\widetilde S(i\infty)$. Therefore,
equation (\ref{S01}) implies (\ref{resu}).

The proof of $(ii)$ is almost identical to that of the last statement of Theorem 2.2.

\medskip
\noindent
{\bf Small norm case.} Now we are in the position to present another case, for which
once again the notion of monodromies around the poles of $A(z)$ can be introduced.
This case is of special importance for further considerations.

For simplicity, it is assumed throughout this subsection that
$\Re\, z_k<\Re\, z_m, \ k<m$. Let us fix a number
$\e<<{\rm max}_{km}|\Re\, z_k-\Re\, z_m|$ and consider
the space of matrix functions $A(z)$ of the form (\ref{rs}) such that the euclidian norm
$|A_k|<\e/2$. If $\e$ is small enough, then $A(z)$ is invertible for $|z-z_k|>\e$,
and therefore, zeros of $\det A$ are localized in the neighborhoods of the poles.
Let us denote them by $z_{ks}^-$:
\beq\label{z-}
\det A(z_{ks}^-)=0,\ \ |z_k-z_{ks}^-|<\e,\ \ s=1,\ldots, \ h_k={\rm rank}\  A_k.
\eeq
Furthermore, for small enough $\e$ a solution of the singular equation
(\ref{K}) for $x_k=(\Re\, z_k+\Re\, z_{k+1})/2, \ k=1,\ldots, n-1,$  can be constructed by
the same iterations (\ref{Ka}) as it was done before for $|x|>>0$.
The corresponding canonical solution $\Psi_k=\Psi_{x_k}$ of (\ref{1}) has poles at
the points $z_m+l, \ l=1,\ldots, \ k\leq m$ and at the points $z_{ms}^--l, \
l=0,\ldots,\ \ m\leq k.$ Then, along lines identical to those used for the proof of
Theorem 2.2, we obtain the following statement.

\begin{theo} There exists $\e$ such that, if $|A_k|<\e$ and satisfy (\ref{res}), then
the corresponding regular singular equation (\ref{1})
has a set of unique meromorphic solutions $\Psi_k,\ k=0,\ldots,n$,
which are holomorphic in the strips $r_{k}+\e<\Re\, z<r_{k+1}+1$, and
grow at most polynomially as $|\Im\, z|\to \infty$,  and are normalized by the condition:
$\lim_{\Im\, z\to -\infty}\Psi_kz^{-K}=1.$

$(i)$ The solutions $\Psi_k$ are asymptotically represented by $Yg_k^{\pm}$,
as $\Im\, z\to \pm\infty$;\, $g_k^-=1$, $g_0^+=g_n^+=1.$

$(ii)$ The local connection matrices $M_k=\Psi_k^{-1}\Psi_{k-1}, \ k=1,\ldots, n,$ have
the form
\beq\label{m250}
M_k=1-{m_k\over e^{2\pi i (z-z_k)}-1}\, ,
\eeq
where $m_k$ are matrices such that
\beq\label{detb0}
(1+m_n)\ldots(1+m_1)=e^{2\pi i K}.
\eeq

$(iii)$ The map $\{A_m\}\longmapsto \{m_k\}$ is a one-to-one correspondence
of the space of matrices $|A_m|<\e$, satisfying
(\ref{res}), and an open neighborhood of the point $(m_k=0)$
of the variety defined by equation (\ref{detb0}).
\end{theo}

\subsection{Mild equations}
In this subsection the previous results are extended to the case of mild differential
equations (\ref{1}) with diagonalizable leading coefficient
\beq\label{mi1}
A=A_0+\sum_{m=1}^n {A_m\over z-z_m}, \ \ A_0^{ij}=\rho_i\delta^{ij}
\eeq
If $\rho_i\neq \rho_j$, then (\ref{1}) has unique formal solution of the form (\ref{Y1}).
The substitution of (\ref{Y1}) into (\ref{1}) gives a set of equations for
$\chi_s$. The first nontrivial equation
\beq
[A_0,\chi_1]={\sum_{m=1}^n A_m}-K
\eeq
defines the diagonal matrix
\beq\label{K11}
K^{ij}=k_i\delta^{ij},\ \ k_i=\sum_{m=1}^n A_m^{ii},
\eeq
and the off-diagonal part of the matrix $\chi_1$. On each step the consecutive equation
defines recurrently  the diagonal entries of $\chi_{s-1}$ and the off-diagonal part
of $\chi_s$.

First, let us consider the case of the {\bf real exponents}.
\begin{theo} Let $A$ be a matrix of the form (\ref{mi1}) with $\rho_i\neq \rho_j,\
\Im \rho_i=0$. Then:

$(A)$ there are unique meromorphic solutions $\Psi_{l},\Psi_{r}$ of equation (\ref{1}),
which are holomorphic in the domains $\Re z<<0$ and $\Re z>>0$, respectively, and which are
asymptotically  represented by $Yg_{l(r)}^\pm,\ g_{l(r)}^-=1,$ as
$\Im z\to \pm\infty$; the matrices $g_{r(l)}= g_{r(l)}^+$ satisfy
the constraints (\ref{T});
\beq\label{T1}
g_{r(l)}^{ii}=1, \ g_{r}^{ij}=0,\ {\rm if}\ \rho_i<\rho_j,\ \
g_{l}^{ij}=0,\ {\rm if} \ \rho_i>\rho_j,
\eeq

$(B)$ the connection matrix $S=(\Psi_r)^{-1}\Psi_l$
has the form
\beq\label{K12}
S(z)=1-\sum_{m=1}^n {S_m\over e^{2\pi i(z-z_m)}-1}, \
S_{\infty}=1+\sum_{m=1}^n\, S_m=g_r^{-1}e^{2\pi i K}g_l\,;
\eeq
\end{theo}
If the case of real exponents $\Im \rho_i=0$ the matrix
$e^{z\ln A_0+z K}$ grows at most polynomially as $|\Im\, z|\to \infty$,
and almost all the results proved above for the regular singular equations
hold. Lemma 2.1 does not require any changes at all. As before, it implies
the existence of meromorphic canonical solutions $\Psi_{r}$ and $\Psi_{l}$ of (\ref{1}).
These solutions are asymptotically represented by $Yg_{l(r)}^{\pm}, \ \Im z\to \pm \infty$.
They are uniquely normalized by the condition $g_{l(r)}^-=1$.
The only difference of mild equations with distinct real exponents and
regular singular equations is that for the first ones
equation $g_{l(r)}=1$ does not hold. The coefficient $(\widetilde A-1)$ in
(\ref{K}) is of the form
\beq\label{diff1}
\widetilde A-1=e^{-z\ln A_0-K\ln z}\,0(z^{-m})\,e^{z\ln A_0+K\ln z}
\eeq
From (\ref{K}) it follows that $\varphi$ asymptotically has
the quasi-triangular form. Then equation (\ref{am3}) implies (\ref{T1}).
The proof of (\ref{K12}) is identical to that of (\ref{S01}).

Let us consider now the {\bf Birkhoff's case} of exponents $\rho_i$ with distinct imaginary parts
of $\ln \rho_i$. Below we assume that the branch of $\ln \rho_i$ is chosen such that
\beq\label{nu}
-\pi< \nu_i=\Im (\ln \rho_i)\leq \pi.
\eeq
\begin{theo} Let $A$ be a matrix of the form (\ref{mi1}) with $\nu_i\neq \nu_j\neq0$. Then:

$(A)$ there are unique meromorphic solutions $\Psi_{l}, \Psi_{r}$ of equation (\ref{1}),
which are holomorphic in the domains $\Re z<<0$ and $\Re z>>0$, respectively, and which are
asymptotically  represented by $Y, \ \Im z\to \pm\infty$;

$(B)$ the connection matrix $S=(\Psi_r)^{-1}\Psi_l$
has the form
\beq\label{K121}
S(z)=S_0-\sum_{m=1}^n {S_m\over e^{2\pi i(z-z_m)}-1},
\eeq
where $S_0$ and $S_{\infty}=1+\sum_{m=1}^n S_m$ satisfy the constraints:
\beq\label{tr}
S_0^{jj}=1,\ \ S_0^{ij}=0,\ {\rm if }\ \nu_i>\nu_j, \ \ \
S_{\infty}^{jj}=e^{2\pi i k_j},\ \ S_{\infty}^{ij}=0,\ {\rm if }\ \nu_i<\nu_j.
\eeq
\end{theo}
The first statement of the theorem is one of the fundamental Birkhoff's results.
Nevertheless, it is instructive to outline its proof via the Riemann-Hilber factorization
problem (\ref{R1}). It clarifies the similarity and the difference of the Birkhoff's case
and the case of  real exponents. The differences are mainly due to the simple fact
that in case $\nu_i\neq\nu_j$ the formal series $Y$ and $Yg$ are asymptotically equal
to each other, as $\Im z\to \pm \infty$, if $g$ is quasi upper- or lower-triangular
matrix , respectively, whose diagonal entries equal $1$. As a result,
the notion of the transfer matrix $g_x$ along the thick path $\Pi_x$ introduced above has
no intrinsic meaning in the Birkhoff's case. It is hidden in the normalization of
$\Psi_{l(r)}$, and to some extend, re-appear in the form of the connection matrix $S$.

As above, the construction of a sectionally holomorphic solution $\Phi_x$ of the
Riemann-Hilbert factorization problem (\ref{R1}) is reduced to a singular
integral equation. Let $\Phi_x$ be a function given by the formula
\beq\label{psia}
\Phi_x=Y_m\phi,\ \ \phi=g+\int_L\varphi(\xi)k(z,\xi)\,d\xi.
\eeq
The function $\Phi_x$ is a solution of the Riemann-Hilbert problem if $\varphi\in H$
is a solution of the singular integral equation
\beq\label{Kg}
(\widetilde A+1)\,\varphi-2(\widetilde A-1)I_{\varphi}=2(\widetilde A-1)g,
\eeq
where $\widetilde A$ is given by (\ref{am}). For regular singular equations and for the case
of mild equations with real exponents a choice of the constant term $g$
in (\ref{psia}) was inessential. It becomes crucial for the case of imaginary exponents.

Our next goal is to show that there exists a unique matrix $g$ whose diagonal entries
equal $g^{ii}=1,$ and such that equation (\ref{Kg})
has a solution $\varphi\in H$ with entries satisfying the conditions
\beq\label{bi1}
|\varphi^{ij}(\xi)|< O(\,|y|^{-m+\kappa})\ e^{\,y\,\nu_{ij}}, \ \nu_{ij}=\nu_i-\nu_j,\ \
y=\Im \xi\to \pm\infty.
\eeq
If a smooth matrix function $\varphi$ satisfies (\ref{bi1}), then the corresponding Cauchy
integrals have the following asymptotics
\beq\label{bi2}
\pm\nu_{ij}>0:\ \ \ \
\left\{
\begin{array}{cll}\ \left|I_{\varphi}^{ij}\right|&<O(\,|y|^{-m+\kappa})\ e^{\,y\,\nu_{ij}},&
y\to \pm\infty,\\
|I_{\varphi}^{ij}-f_{\varphi}^{ij}|&<O(\,|y|^{-m+\kappa})\ e^{\,y\,\nu_{ij}},&
y\to \mp\infty, \end{array}\right.
\eeq
where
\beq\label{bi3}
\pm \nu_{ij}>0: \ \ f_{\varphi}^{ij}=-{1\over 2}\int_L(\tan(\pi y)\pm 1)\,
\varphi^{ij}(\xi)d\xi.
\eeq
The proof of the second inequality in (\ref{bi2})
is almost identical to that of (\ref{am2}). The first inequality can be obtained by
similar arguments (see also formula () in \cite{mus}).

Self-consistence of equation (\ref{Kg}) and
the conditions (\ref{bi2}) implies
\beq\label{Kg1}
g=1-f_{\varphi}\,,
\eeq
where $f_{\varphi}$ is an off-diagonal matrix given by (\ref{bi3}). Equations (\ref{Kg})
and (\ref{Kg1}) can be seen as a system of equations for unknown $\varphi(\xi)$ and $g$.
This system for a large $|x|$ can be solved by iterations. For that we take $\varphi_0=0$
and define $\varphi_n$ recurrently by the equation
\beq\label{Kar}
(\widetilde A+1)\,\varphi_{n+1}=
2(\widetilde A-1)(1+I_{\varphi_n}-f_{\varphi_{n}}),
\eeq
From (\ref{bi2}) it follows that if $\varphi_n$ satisfies (\ref{bi1}),
then $\varphi_{n+1}$ satisfies the same conditions, as well.
The sequences $g_n=1-f_{\varphi_n}, \ \varphi_n$ converge and define $g$ and
a solution $\varphi$ of (\ref{Kg}), which  satisfies (\ref{bi2}).

From (\ref{bi2}) it follows that if $\phi$ and $g$ are solutions of (\ref{Kg}) and
(\ref{Kg1}), then the off-diagonal entries of the matrix function $\Phi$ given by
(\ref{psia}) have the asymptotic
\beq\label{psib}
|\phi^{ij}(z)|< O(\,|z|^{-m+\kappa})\ \left|\left(\rho_j/\rho_i\right)^z\right|,\ \
\Im z\to \pm\infty.
\eeq
on both the edges of $\Pi_{x,\e}$.
For the diagonal elements of $\phi$ we have the same asymptotic
as that proven above for the case of regular singular equations, i.e.
\beq\label{psic}
|\phi^{jj}(z)-v_j^{\pm}|< O(\,|z|^{-m+\kappa+1}),\ \
\Im z\to \pm\infty,
\eeq
where
\beq\label{am3a}
v_j^{\pm}=1-{1\over 2}\int_L(\tan(\pi i(\xi-x))\pm 1)\, \varphi^{jj}(\xi)\, d\xi.
\eeq
The same arguments as used above in the section 2.1, show that, if there exists a
sectionally meromorphic solution $\Phi_x$ of the Riemann-Hilbert problem (\ref{R1}),
then it is unique. Therefore, (\ref{psib},\ref{psic}) imply the following statement.
\begin{lem} For a generic $A$, such that $ind_x \,A=0$ there exists a unique holomorphic
solution $\Phi_x$ of the Riemann-Hilbert Problem (\ref{R1}) asymptotically represented by
$Y$, as $\Im z\to -\infty$, and $Yv_x$, as $\Im z\to \infty$, where $v_x$ is a diagonal matrix.
\end{lem}
For large $|x|$ the solutions $\Psi_x$ of equation (\ref{1}) corresponding to
$\Phi_x$ are $x$-independent for $x>>0$ and $x<<0$ and can be identified with the Birkhoff's
solutions $\Psi_r$ and $\Psi_l$, respectively. Indeed, for a large $|x|$
the functions $\varphi^{ii}$ are uniformly bounded by $O(|x|^{-m+\kappa})$. Therefore,
(\ref{psic}) implies $v_{l(r)}=1.$ The first statement of the theorem is proved.

From (\ref{nu}) it follows that the connection matrix $S$ considered as a function of the
variable $w=e^{2\pi iz}$ has holomorphic extension at the points $w=0,w=\infty$. Therefore,
it is a rational function of $w$ having poles at $w_m=w(z_m)$. Hence, it has the form
(\ref{K12}). Its evaluations at $w=0$ and $w=\infty$ are quasi triangular matrices for
obvious reasons. The proof of the theorem is completed.

\bigskip
\noindent{\bf Local monodromies} for mild equations can be introduced for the same
three cases considered above in the section 2.2. Namely, for the cases
of {\it special, unitary} and {\it small norm} coefficients. The form of the local
monodromy matrices in the case of mild equations with real exponents
was described in the Introduction. Extensions of all the other results
of the section 2.2 for the case of mild equations are straightforward.
For example let us consider the special mild equations with imaginary exponents,
satisfying the Birkhoff's condition.

\begin{theo} $(i)$ For a generic matrix $A$ of the form
\beq\label{A1m}
A(z)=A_0\left(1+{a_nb_n^T\over z-z_n}\right)\cdots
\left(1+{a_1b_1^T\over z-z_1}\right)\,
\eeq
where
$$ (a)\ A_0^{ij}=\rho_i\delta^{ij},\ \nu_i\neq\nu_j, \ \nu_i=\Im (\ln \rho_i),\ \
(b) \ b_k^Ta_k=0,\ \ (c)\
\Re z_k<\Re z_m, \ k<m,
$$
equation (\ref{1}) has a set of unique meromorphic solutions $\Psi_k,\ k=0,\ldots,n$,
which are holomorphic in the strips $r_{k}<\Re\, z<r_{k+1}+1$, and
asymptotically represented by $Yv_k^{\pm}$,\ as $\Im\, z\to \pm\infty$,
where $v_k^-=1$, and $v_k^+$ is a diagonal matrix

$(ii)$ The local connection matrices $M_k=\Psi_k^{-1}\Psi_{k-1}, \ k=1,\ldots, n,$
have the form
\beq\label{m25m}
M_k=m_{k0}-{\a_k\b_k^T\over e^{2\pi i (z-z_k)}-1}\ \
\eeq
where $(\a_k,\b_k)$ are pairs of orthogonal vectors,
considered modulo transformations (\ref{gauge}), and $m_{k0}$ is a quasi lower-triangular
matrix such that $M_k(i\infty)$ is a quasi upper-triangular matrix, i.e.
\beq\label{trm}
m_{k0}^{jj}=1,\ \ m_{k0}^{ij}=0,\ {\rm if }\ \nu_i>\nu_j, \ \ \
m_{k0}^{ij}=-\a_k^i\b_k^j,\ {\rm if }\ \nu_i<\nu_j,
\eeq
$(iii)$ The map of pairs of orthogonal vectors $\{a_m,b_m\}\longmapsto \{\a_k,\b_k\}$,
considered modulo transformation (\ref{gauge}), is a one-to-one correspondence of open sets.
\end{theo}
In the small norm case the local connection matrix $M_k$ is described in similar terms.
Namely, it has the form
\beq\label{m25m1}
M_k=m_{k0}-{m_{k1}\over e^{2\pi i (z-z_k)}-1}\, ,\ \
\eeq
where $m_{k0}$ is quasi lower-triangular matrix and $m_{k0}+m_{k1}$ is a quasi
upper-triangular matrix. The discrete analog of the local monodromy matrix is defined
as their ratio
\beq\label{muka}
\mu_k=1+m_{k1}m_{k0}^{-1}
\eeq
Note, that a generic matrix  has a unique factorization as the product of
lower- and upper-triangular matrices. Therefore, equation (\ref{muka}) implies
that $\mu_k$ uniquely defines the corresponding pair of matrices $m_{k0},m_{k1}$,
and, consequently, the local and global connection matrices.

\section{The inverse monodromy problem and isomonodromy transformations}

In this section we consider a map inverse to the direct monodromy map
\beq\label{dirmap}
\{z_m, A_m\}\longmapsto \{w_m,S_m\},\ \ w_m=w(z_m)=e^{2\pi i z_j}\,.
\eeq
For any fixed diagonalizable matrix $A_0$ the characterization of equations (\ref{1})
having the the same monodromy data is identical to that given by Birhhoff for the case
of imaginary exponents.

\begin{lem} Rational functions $A(z)$ and $A'(z)$
of the form (\ref{L}) under the map (\ref{dirmap}) correspond to the same
connection matrix $S(z)$ if and only if there exists a rational matrix function
$R(z)$ such that
\beq\label{iR}
A'(z)=R(z+1)A(z)R^{-1}(z),\ \ R(\infty)=1.
\eeq
\end{lem}
{\it Proof.} Let $\Psi_{l(r)}$ and  $\Psi'_{l(r)}$ be canonical meromorphic solutions of
equation (\ref{1}) corresponding to $A(z)$ and $A'(z)$, respectively.
If $\Psi_r^{-1}\Psi_l=(\Psi_r')^{-1}\Psi'_l$, then
\beq\label{iR1}
R=\Psi'_l\Psi_l^{-1}=\Psi'_r\Psi_r^{-1}.
\eeq
By definition of the canonical solutions, the matrix function $R$ is holomorphic
for large $|z|$. Moreover, if $A_0=A_0', K=K'$, then $R\to 1, |z|\to \infty$.
Hence, $R$ has only finite number of poles, and therefore, is a rational function of
the variable $z$.

Let $\A_{D}$ be the subspace of the space $\A$ of matrix functions of the form (\ref{L})
having fixed determinant
\beq\label{delta}
A\in \A_{D}\subset \A: \ \
\det A(z)=D(z)={\prod_{\a=1}^N (z-\z_{\a})\over \prod_{m=1}^n(z-z_m)^{h_m}},\ \
h_m= {\rm rk}\ A_m.
\eeq
Note, that the constraint (\ref{trace}) is equivalent to the condition
\beq\label{D1}
{\rm tr}\  K=0\longleftrightarrow \sum_{\a}\z_{\a}=\sum_m h_m z_m.
\eeq
\begin{lem} If the zeros $\z_{\a}$ are not congruent, i.e. $\z_{\a}-\z_{\b}\notin Z$, then
the monodromy correspondence (\ref{dirmap}) restricted to $\A_D$ is
injective.
\end{lem}
{\it Proof.} Let $\A\in \A_D$ be a matrix whose poles and zeros of the determinant are
not congruent pairwise . Suppose that there exists a rational matrix function $R$ that
equals $1$ at the infinity, i.e. $R=1+O(z^{-1})$, and such that the matrix
$A'$ defined by (\ref{iR1}) has the same determinant, i.e. $A'\in \A_D$.
Then the equation $R(z+1)=A'(z)R(z)A^{-1}(z)$ implies that $R$
has poles of {\it constant} ranks at the points $\z_{\a}+l$ and $z_m+l$,
where $l\in Z_+$ is a positive integer. The matrix $R$ is regular at the infinity.
Therefore, it should be regular everywhere. That implies $R=1$.

Let us call rational functions $D$ and $D'$ equivalent if sets of
their poles $z_m, z_m'$ and zeros $\z_{\a},\z_{\a}'$ are congruent to each other, i.e
$z_m-z_m'\in Z, \ \ \z_{\a}-\z_{\a}'\in Z$, and satisfy the relation (\ref{D1}).

\begin{lem} For each pair of equivalent rational functions $D$ and $D'$ there exists
a unique isomonodromy birational transformation
\beq\label{Trans}
T_D^{D'}:\A_D\longmapsto \A_{D'}
\eeq
\end{lem}
{\it Proof.} The construction of the isomonodromy transformations $T_D^{D'}$ is analogous
to that proposed in \cite{bor} for the case of polynomial coefficients $\widetilde A$.
To begin with, we introduce two types of elementary transformations.
They are birational and defined on open sets of the corresponding spaces.
An elementary isomonodromy transformation of the first type is defined by
a pair $z_k, \z_{\a}$ and the eigenvector of $A_k=\res_{z_k} A$, corresponding to
a non-zero eigenvalue $\lambda$,
\beq\label{eigen}
q^T A_k=\lambda q^T\neq 0.
\eeq
Consider the matrix
\beq\label{i20}
R=1+{pq^T\over z-z_k},
\eeq
where $p$ is the null-vector of $A(\z_{\a})$ normalized so that
\beq
\label{i21}
(q^Tp)=z_k-\z_{\a}, \ \ A(\z_{\a})p=0.
\eeq
{\it Remark.} If $z_k\neq \z_{\a}$, then the matrix $R$ is defined only on an open set
of $\A_D$, where the product $(q^Tp)$ of the corresponding eigenvectors is non-zero.

Equation (\ref{i21}) implies
\beq\label{i22}
R^{-1}=1-{pq^T\over z-\z_{\a}}.
\eeq
Furthermore, from the second equation (\ref{i21}) it follows, that the matrix $A'$
given by (\ref{iR}) is regular at $\z_{\a}$. The matrix $A'$ has a pole of rank 1
at $z_k-1$. The rank of its residue at $z_k$ is
equal to the rank of the matrix $A_kR^{-1}(z_k)$. The left null-space of the last
matrix contains the null-space of $A_k$ and the vector $q^T$. Hence, the residue of
$A'$ at $z_k$ has rank $h_k-1$. In the same way, choosing another zero $\z_{\a_2}$ of $D$
and the eigenvector of $A'_k=\res_{z_k}A'$ corresponding to a non-zero eigenvalue, we get
a matrix function $A''$ with a pole at $z_k$ of rank $h_k-2$.
Further iterations give a matrix $T_k^{\a_1,\ldots,\a_{h_k}}(A)$, which is regular at $z_k$
and has a pole of rank $h_k$ at $z_k-1$.

As follows from Lemma 3.2, the isomonodromy transformation $T_k^{\a_1,\ldots,\a_{h_k}}$
is uniquely defined by the choice of a pole $z_k$ and a subset of $h_k$ zeros
$\z_{\a_s}$ of $D$. These transformations are analogs of isomonodromy transformations
introduced in \cite{bor} for the case of polynomial $A(z)$.

An elementary isomonodromy transformation of the second type is defined by a pair of
zeros $\z_{\a}$ and $\z_{\b}$ of $D$. The corresponding matrix $R=R_{\a,\b}$
is given by the formula
\beq\label{i23}
R_{\a,\b}=1+{p_{\a}q_{\b}^T\over z-z_{\b}-1},
\eeq
where $p_{\a}$ and $q_{\b}$ are vectors defined by the equations
\beq\label{i24}
(i)\  A(\z_{a})p_{\a}=0;\ (ii)\  q_{\b}^TA(\z_{\b})=0; \
(iii)\left(q_{\b}^Tp_{\a}\right)=\z_{\b}-\z_{\a}+1.
\eeq
From (\ref{i24} $iii$) it follows that $R^{-1}_{\a,\b}=1-p_{\a}q_{\b}^T/(z-z_{\a})$.
Then equations (\ref{i24} $i, ii$) imply that the matrix
\beq\label{i25}
T^{\a|\b}(A)=R_{\a,\b}^{-1}(z+1)A(z)R_{\a,\b}^{-1}(z)=
\left(1+{p_{\a}q_{\b}^T\over z-z_{\b}}\right)A(z)
\left(1+{p_{\a}q_{\b}^T\over z-z_{\a}}\right)
\eeq
is regular and non-degenerate at $\z_{\a}$ and $\z_{\b}$. It has the same set of poles
as $A$. The zeros of its determinant are $\z_{\a}-1,\ \z_{\b}+1$ and $\z_{\g},
\ \g\neq \a,\b$.

The transformation $T_D^{D'}$ can be obtained as the composition of elementary
isomonodromy transformations. Indeed, if $D$ and $D'$ are equivalent, then
the poles of $D$ can be shifted to the poles of $D'$ by elementary transfomations
(or their inverse) of the first type. After that $N-1$ zeros can be shifted to
$N-1$ zeros of $D'$ by transformations of the second type. Then, equation (\ref{D1})
defines a unique position of the last zero. The lemma is proved.

Now we are ready to present the main result of this section.
\begin{theo} Let $A_0^{ij}=\rho_i\delta^{ij}$ and $K$ be diagonal matrices and let
$S(w)$ be a rational matrix function of the variable $w=e^{2\pi i z}$
having the form:
(a) (\ref{S01}), if $A_0=1$, (b) (\ref{T1},\ref{K12}), if $\Im \rho_i=0$; (c) (\ref{K121},\ref{tr})
if $\Im \ln \rho_i\neq\Im \ln \rho_j\neq 0$. Then, for each $S$ in general position
and for each set of branches $z_k,\z_{\a}$ of the logarithms of poles and zeros of
$\det S(w)$, there exists a unique rational matrix function
$A(z)$ of the form (\ref{L}) such that $S(z)$ is the connection matrix of the corresponding
difference equation (\ref{1}) and $\det A(\z_{\a})=0.$
\end{theo}
{\it Proof.} It has been already proven that if $A(z)$ exists for one set of
$z_k,\z_{\a}$, then in general position it exists and is unique for any equivalent set.
Therefore, for the proof of the theorem it is enough to construct one
equation (\ref{1}) for which $S$ is the connection matrix.

Let us fix a real number $x$ such that on the line $L:\Re z=x$ the matrix $S(z)$ is
regular and invertible. We denote the half planes $\Re z<x$ and $\Re z>x$
by $D_l$ and $D_r$, respectively. Consider the following factorization problem

\medskip
\noindent{\bf Problem III.} {\it For a given $S$ find invertible matrix
functions $\X_l(z)$ and $\X_r(z)$, which are holomorphic and bounded
inside the domains $\D_l, D_r$, respectively, continuous up to the boundaries, and
such that the functions $\Psi_{l(r)}=\X_{l(r)}e^{z\ln A_0+K\ln z}$
satisfy the equation }
\beq\label{RIII}
\Psi_l(\xi)=\Psi_r(\xi)S(\xi),\ \ \xi\in L\,.
\eeq
\begin{lem} For a generic matrix $S$ the Problem III has a solution which is unique up
to the normalization $\X'_{l(r)}=g\X_{l(r)}$.
\end{lem}
{\it Proof.} Consider  functions $\X_{l(r)}$ defined in each of the half-planes by
the Cauchy integral
\beq\label{bX1}
\X(z)=1+{1\over 2\pi i}\int_{L} {\chi(\xi)d\xi\over \xi-z}\, .
\eeq
Equation (\ref{RIII}) is equivalent to the equation
\beq\label{Kx1}
{1\over 2}\chi(\xi)(\M(\xi)+1)-{1\over 2\pi i}I_{\chi}(\xi)(\M(\xi)-1)=(\M(\xi)-1),
\eeq
where $\M=Y_0SY_0^{-1},\ \ Y_0=e^{z\ln A_0+K\ln z}$. If $S$ has the form $(a)$ or $(c)$,
then at the infinity  $\M$ exponentially tends to $1$, and for a generic $S$ equation
(\ref{Kx1}) has a unique solution. In the case $(b)$
of the mild equations with real exponents, the coefficient $\M$ has no limit
at the infinity and the fundamental results of the theory of singular integral equations can
not be applied  directly.

The following slight modification of the Problem III allows us to prove the lemma for
the case $(b)$. Consider  functions $\X'_{l(r)}$ given by the Cauchy integral (\ref{bX1}) over the line
$\xi\in L': Arg(\xi-x)=\pi/2 + \e,\, \e >0$. If $\chi(\xi), \xi\in L',$ is a solution of
equation (\ref{Kx1}) on $L'$ with the coefficient $\M'=Y_0g_rSY_0^{-1}$, then the boundary
values of the functions $\Psi'=\X'_lY_0$ and $\X'_rY_0$ on $L'$ satisfy the equation
\beq\label{RIV}
\Psi'_l(\xi)=\Psi'_r(\xi)g_rS(\xi),\ \ \xi\in L'\,.
\eeq
From (\ref{T1}) it follows that $\M'$ along $L'$ exponentially tends
to the identity matrix. Therefore, a solution $\chi$ of the corresponding equation (\ref{Kx1}) on $L'$ exists
and is unique. It defines a unique solution of the factorization problem (\ref{RIV}).
The equation (\ref{RIV}) can be used for meromorphic extension of the functions
$\Psi'_{l(r)}$, which are originally defined in the half-planes separated by $L'$.
If $\e>0$ is small enough, then $S$  is regular and
invertible in the sectors between $L$ and $L'$.
Hence, the extensions of the functions $\Psi'_l$ and $\Psi'_r$
are holomorphic in the domains $\D_l$ and $D_r$, respectively. Therefore,
the functions $\Psi_l=\Psi_l'$ and $\Psi_r=\Psi'_rg_r$ are solutions of the factorization
problem (\ref{RIII}). The lemma is proved.

Let $\Psi_{l(r)}$ be a solution of the factorization problem (\ref{RIII}).
Then the function
\beq\label{Af}
A(z)=\Psi_l(z+1)\Psi_l^{-1}(z)=\Psi_r(z+1)\Psi_r^{-1}(z)
\eeq
is holomorphic in the domains $\Re z< x-1$ and $\Re z > x$. It tends to $A_0$ as $z\to \infty$.
Inside the strip $\Pi_{x-1}$
the poles of $A$ and $A^{-1}$ coincide with the zeros and the poles of
$S$ and $S^{-1}$. Therefore, $A(z)$ has the form (\ref{L}), where $x-1<\Re z_m<x$.
The theorem is thus proven.

\section{Continuous limit}

Our next goal is to show that in the continuous limit $h\to 0$
the canonical meromorphic solutions $\Psi_x$ of difference equation (\ref{1e})
converge to solutions of differential equation (\ref{1dif}).

The construction of canonical meromorphic solution $\Psi_x$ of (\ref{1e}), which is
holomorphic in the strip $\Pi_x^h: x<\Re z< x+h,$ requires only slight changes
of the formulas used above.
As before, a sectional holomorphic solution $\Phi_x$ of the factorization problem
\beq\label{R1h}
\Phi_x^+(\xi+h)=(1+hA(\xi))\Phi_x^-(\xi),\ \ \xi=x+iy\,
\eeq
can be represented with the help of the Cauchy type integral
\beq\label{psih}
\Phi_x=Y_0\phi,\ \ \phi=1+\int_L\varphi(\xi)k_h(z,\xi)\,d\xi,\
k_h=k(h^{-1}z,h^{-1}\xi),
\eeq
where $k(z,\xi)$ is given by (\ref{ker}) and $Y_0=e^{z\ln(1+hA_0)+hK\ln z}$.
The residue of $k_h$ at $z=\xi$ equals $h$, therefore the boundary values of $\phi$ are
\beq\label{psi1h}
\phi^-(\xi)=
-{h\varphi(\xi)\over 2}+
1+I_{\varphi}(\xi),\ \
\phi^+(\xi+1)={h\varphi(\xi)\over 2}+1+I_{\varphi}(\xi) \,,
\eeq
where $I_{\varphi}$ is the principal value of the corresponding integral.
The singular integral equation for $\varphi$ which is equivalent to (\ref{R1h}) now
takes the form
\beq\label{Kh}
(2+h\widetilde A)\,\varphi-2\widetilde A I_{\varphi}=2\widetilde A\,,
\eeq
where $\widetilde A=Y_0(\xi+1)^{-1}A(\xi)Y_0(\xi)$. If $|\,x-\Re z_k|>Ch,$
then equation (\ref{Kh}) can be solved by iterations. As before, the corresponding solution
$\Psi_x$ is $x$-independent in the intervals $\Re z_k+Ch<x<\Re z_{k+1}-Ch$. Hence,
we conclude that: for any $\e>0$, and any rational function $A(z)$
of the form (\ref{L}) there exist $h_0$ such that equation (\ref{1e}) for $h<h_0$
has canonical meromorphic solutions $\Psi_k$, which are holomorphic in the strips
$z\in \D_k:\Re z_k+\e<\Re z<\Re z_{k+1}-\e$.

The existence of $\Psi_k$ implies that for each $A$ of the form (\ref{L})
the local monodromy matrices $\mu_k$ are well-defined for sufficiently small $h$.
Hence, we may consider their continuous limits.

\begin{theo} In the limit $h\to 0$:

$(A)$ the canonical solution $\Psi_k$ of difference equation
(\ref{1e}) uniformly in $\D_k$ converges to a solution $\hat \Psi_k$ of differential equation
(\ref{1dif}), which is holomorphic in $\D_k$;

$(B)$ the local monodromy matrix (\ref{mu}) converges to the monodromy of $\hat\Psi_k$
along the closed path from $z=-i\infty$ and goes around the pole $z_k$;

$(C)$ the upper- and lower-triangular matrices $(g_{r}, g_l)$ and  $(S_0,S_{\infty})$
defined in (\ref{T1}) and (\ref{tr}), respectively, for the cases of real and imaginary
exponents, converge to the Stokes' matrices of differential equation (\ref{1dif}).
\end{theo}
The first statement of the theorem follows from a simple observation, that in the continuous
limit the singular integral equation for solutions of the Riemann-Hilbert factorization
problem becomes just differential equation (\ref{1dif}).
It is easy to check that
\beq\label{kerh}
k_h(z,\xi)=\left\{\begin{array}{rl} 1+O(h), &
z-\xi > h\ln h, \ \xi>h\ln h,\\
O(h), &
\xi-z> h\ln h, \ {\rm or}\ \xi<h\ln h,\end{array}
\right. z>h\ln h
\eeq
Similar equations are valid for $z<h\ln h$. In both the cases we have
\beq\label{h10}
I_{\varphi}(z)=\int_0^{z}\varphi(\xi)d\xi +O(h)
\eeq
From (\ref{Kh}, \ref{h10}) it follows that the function $\psi=1+I_{\varphi}$
satisfies the relation
\beq\label{Khh}
{d\psi\over dz}=A(z)\psi(z)+O(h).
\eeq
On the line $L_x:\Re z=x$ the function $\Phi_x$ equals $\psi+0(h)$. Therefore, $\Phi_x$
does converge to $\hat \Psi_k$ on $L_x$. For the cases of regular singular
equations and mild equations with real exponents the convergence is uniform on
$\D_k$. For the case of imaginary exponents the convergence becomes uniform only
for the special choice of constant term $g$ in the integral
representation for $\Phi_x$, which in (\ref{psih}) was set $g=1$
(compare with (\ref{psia})).

The second and the third statements of the theorem are direct corollaries of $(A)$ and
of the definition of the local monodromy matrices $\mu_k$ and the matrices
$(g_{r}, g_{\,l})$ and  $(S_0,S_{\infty})$.

\section {Difference equations on elliptic curves}

In this section we construct direct and inverse monodromy maps for difference equations
on an elliptic curve.

Let $\G$ be the elliptic curve with periods $(2\omega_1, 2\omega_2),\
\Im (\omega_2/\omega_1)>0$. Consider the equation
\beq\label{el1}
\Psi(z+h)=A(z)\Psi(z),
\eeq
where $A(z)$ is a meromorphic $(r\times r)$ matrix function with simple poles,
which satisfies the following monodromy properties
\beq\label{elmon}
A(z+2\omega_{\a})=B_{\a}A(z)B_{\a}^{-1},\ \ B_{\a}\in SL_r
\eeq
The matrix $A(z)$ can be seen as a meromorphic section of the vector bundle $Hom (\V,\V)$,
where $\V$ is the holomorphic vector bundle on $\G$ defined by a pair of commuting
matrices $B_{\a}$. Throughout this section it is always assumed that $B_{\a}$
are diagonalizable. Equation (\ref{el1}) is invariant under
the gauge transformation $A'=GAG^{-1}$. Therefore, if $B_{\a}$ are diagonalizable, then
we may assume without loss of generality, that $B_{\a}$ are diagonal.
Furthermore, if $G$ is a diagonal matrix, then equation (\ref{el1}) is invariant under
the transformation
\beq\label{elg}
\Psi'=G^z\Psi,\ \ A'=G^{z+h} A(z) G^{-z}, G^{ij}=G^i\delta^{ij}
\eeq
The matrix $A'$ has the following monodromy properties
\beq\label{elmon1}
A'(z+2\omega_{\a})=B'_{\a}A'(z)\left(B_{\a}'\right)^{-1},\ \
B'_{\a}=G^{2\omega_{\a}}B_{\a}.
\eeq
Therefore, if $B_{\a}$ are diagonalizable, then we may assume without loss of generality,
that
\beq\label{B}
B_1^{lj}=\delta^{lj},\ \ \ B_{2}^{lj}=e^{\pi i q_j/\omega_1 } \delta^{ij}.
\eeq
Below we assume that $q_i\neq q_j$.
Entries of the matrix $A$ can be expressed in terms of the standard
Jacobi theta-function: $\theta_3(z)=\theta_3(z|\tau),\ \ \tau=\omega_2/\omega_1$.
Let us define the function $\s$ by the formula
\beq\label{sig}
\s(z)=\s(z|2\omega_1,2\omega_2)=\theta_3(z/2\omega_1|\omega_2/\omega_1).
\eeq
The monodromy properties of $\theta_3$ imply
\beq\label{sig1}
\s(z+2\omega_{1})=\s(z), \ \ \s(z+2\omega_{2})=-\s(z)e^{-\pi iz/\omega_1}.
\eeq
The function $\s$ is an odd function $\s(z)=-\s(-z)$.
From (\ref{sig1}) it follows that the entries of $A$ satisfying (\ref{elmon}, \ref{B})
can be uniquely represented in the form
\begin{eqnarray}\label{elA}
A^{ii}&=&\rho_i+\sum_{m=1}^{n} A_m^i \widetilde\z(z-z_m),\ \ \ \sum_m A_m^i=0,
\nonumber\\
A^{ij}&=&\sum_{m=1}^n A_m^{ij}\  {\s(z-q_i+q_j-z_m)\over\s(z-z_m)},\ \ i\neq j,
\end{eqnarray}
where $\widetilde\z=\p_z(\ln \s)$, and $z_m\in C$ are the poles of $A(z)$ in the
fundamental domain
\beq\label{dom}
0<r(z_m)<1,\ \ 0<u(z_m)<1
\eeq
of $C/\Lambda,\ \ \Lambda=\{2n\omega_1,2m\omega_2\}$.
Here and below we will use the notation $r(z)$ and $u(z)$ for real coordinates
$z=2r\omega_1+2u\omega_2$ of $z\in C$ with respect to the basis $2\omega_{\a}$,
\beq\label{z}
r(z)={z\bar\omega_2-\bar z\omega_2
\over 2(\omega_1\bar \omega_2-\bar\omega_1\omega_2)},\
u(z)={z\bar\omega_1-\bar z\omega_1
\over 2(\omega_2\bar \omega_1-\bar\omega_2\omega_1)},
\eeq
Throughout this section it is assumed that the poles $z_m$
of $A$ are non congruent (mod $h$), i.e.
$h^{-1}(z_m-z_k)\notin Z$.

Our goal is to construct canonical meromorphic solutions of equation (\ref{el1})
with the coefficients of the form (\ref{elA}). As before, this problem is reduced to
a proper Riemann-Hilbert factorization problem. For definiteness we assume that the step
$h$ of the difference equation satisfies the condition
\beq\label{zh}
0<r(h)<1.
\eeq
Let us fix a real number $x$ and consider the following problem in the strip
$z\in \Pi_x: \ x\leq r(z)\leq x+r(h)$.

\medskip
\noindent{\bf Problem IV.}
{\it Find in the strip $\Pi_x$ a continuous matrix function $\Phi(z)$, which is
meromorphic inside $\Pi_x$, and whose boundary values on two sides of the strip satisfy
the equation}
\beq\label{E1}
\Phi^+(\xi+h)=A(\xi)\Phi^-(\xi),\ \ r(\xi)=x.
\eeq
The index of the problem is given by the integral
\beq\label{ind0}
ind_{x}(A)=\int_{L_x}d\ln \det A, \ \xi\in L_x: r(\xi)=x,
\eeq
\begin{lem} For a generic $A(z)$, such that $ind_x(A)=0$, there exists a non-degenerate
holomorphic solution $\Phi_x$ of the problem (\ref{E1}) having the following
monodromy property
\beq\label{mm}
\Phi_x(z+2\omega_2)=e^{\pi i \hat q/\omega_1}\Phi_x(z)e^{-2\pi i \hat s},
\eeq
where $\hat q$ is the diagonal matrix defining the monodromy property
(\ref{elmon},\ref{B}) of $A$, and $\hat s$ is
a diagonal matrix $\hat s^{ij}=s^i\delta^{ij}$.
The solution $\Phi_x$ is unique up to the transformation $\Phi_x'=\Phi_xF$,
where $F$ is diagonal.
\end{lem}
{\it Proof.} The lemma can be proved by methods of algebraic geometry. Indeed,
let us define an action of the lattice $\Lambda_h$ span by $h$ and $2\omega_2$
on $(z,f)\in C\times C^r$ as follows:
\beq\label{ac}
(z,f)\to (z+h,A(z)f),\ \ (z,f)\to (z+2\omega_2, B_2f),\ \ B_2=e^{\pi i \hat q/\omega_1}.
\eeq
Then the factor-space $C\times C^r/\Lambda_h$ is a vector-bundle $\V$ on the elliptic curve
$\G_h$ with periods $(h,2\omega_2)$. From (\ref{ind0}) it follows that
the determinant bundle of $\V$ is of degree zero, $c_1(\V)=0.$ According to \cite{Nar},
for a generic zero degree vector bundle on an algebraic curve there exists
a flat holomorphic connection. A basis of horizontal sections of such
connection can be identified with a holomorphic matrix function $\Phi'$ satisfying the
relations $\Phi'(z+h)=A(z)\Phi'(z)V_1,\ \  \Phi'(z+2\omega_2)=B_2\Phi'(z)V_2,$
where $V_1,V_2$ is a pair of commuting matrices.  The change of the basis of
horizontal sections corresponds to the transformation
$\Phi'\to \Phi g, \ V_i\to g^{-1}V_i g$. Therefore, in the
general position when $V_i$ are diagonalizable, we may assume, without loss of generality,
that $V_i$ are diagonal. Now we can define a holomorphic solution of the
boundary problem (\ref{E1}) as follows $\Phi_x=\Phi'V_1^{-z/h}$. It satisfies
the monodromy relation (\ref{mm}), where $e^{-2\pi \hat s/h}=V_2 V_1^{-2\omega_2/h}$.
We call $\Phi_x$ the {\it Bloch solution} of the factorization problem (\ref{E1}).
In the general position we may assume that $s_i\neq s_j$.

Suppose, that there are two Bloch solutions $\Phi_x$ and $\Phi_x'$ of the
factorization problem (\ref{E1}).
From (\ref{ind0}) it follows that $\Phi_x$ is non-degenerate in $\Pi_x$.
Therefore, the entries of the matrix function $F=\Phi_x^{-1}\Phi'_x$
are holomorphic matrix functions satisfying the relations
\beq\label{eF}
F^{lj}(z+h)=F(z),\ \ F^{lj}(z+2\omega_2)=F^{lj}(z)e^{2\pi i(s_l-s_j')/h}
\eeq
Equations (\ref{eF}) imply that $F^{ij}=0$, if $s_i\neq s_j$. Indeed, consider the
function
\beq\label{tht}
\hat F^{ij}=F^{ij}\s_{h}(z+s_i-s'_j)/\s_{h}(z),
\eeq
where $\s_{h}$ is the function given by the formula (\ref{sig1}) for the $\G_h$, i.e.
\beq\label{thet1}
\s_{h}(z)=\s(z|h,2\omega_2).
\eeq
From (\ref{eF}) it follows, that $\hat F^{ij}$ is an
elliptic function on $\G_h$ with one simple pole at $z=0$. There is no such a non-trivial
function. Hence, $s_i=s_i'$ and $F^{ij}=0, i\neq j$, and the Lemma is thus proven.

Now we are ready to define the direct monodromy map for difference equations (\ref{el1})
with coefficients $A$ of the form (\ref{elA}). As before, a holomorphic solution $\Phi_x$
of the boundary problem (\ref{E1}) defines a meromorphic solution $\Psi_x(z)$ of
(\ref{el1}). From (\ref{mm}) it follows that it satisfies the Bloch relation
(\ref{bloch}).

The matrix $A$ has period $2\omega_1$. That implies
\beq\label{eee}
\Phi_{x+1}(z+2\omega_1)=\Phi_x(z),\ \ z\in \Pi_x\,.
\eeq
Therefore, the matrix $\Psi_x(z-2\omega_1)$ is the Bloch solution of (\ref{el1}), which
is holomorphic in the strip $\Pi_{x+1}$. Let us consider the connection matrix of
two Bloch solutions
\beq\label{ss1}
S_x(z)=\Psi_{x}^{-1}(z-2\omega_1)\Psi_x(z).
\eeq
For obvious reason the matrix $S_x$ is $h$-periodic. Let us show that it has the
following monodromy properties
\beq\label{mS}
S(z+h)=S(z),\ \ S(z+2\omega_2)=e^{2\pi i\hat s/h}S(z)e^{-2\pi i\hat s/h},
\eeq
where $\hat s$ is the diagonal matrix defined by the monodromy properties (\ref{mm})
of $\Phi_x$.

By definition the connection matrix $S_x$ depends on the choice of $x$. Let us fix $x=0$,
and denote $S_{x=0}(z)$ by $S(z)$.

\begin{theo}
In the general position the entries of the monodromy matrix $S(z)$ have the form
\begin{eqnarray}\label{eS}
S^{\,ii}&=&S_0^{\,i}+\sum_{m=1}^n S^{\, i}_m \zeta_{h}(z-z_m),\ \sum_{m=1}^n S^{\,i}_m=0,
\nonumber \\
S^{ij}&=&\sum_{m=1}^n S_m^{\,ij}\ {\s_{h}(z-s_i+s_j-z_m)\over \s_{h}(z-z_m)}\,, \ i\neq j.
\end{eqnarray}
where $\zeta_{h}=\p_z\ln \s_{h}$, and $\s_h$ are given by (\ref{thet1}).
\end{theo}
Recall, that $z_m$ are the poles of $A(z)$ in the fundamental domain (\ref{dom})
of $C/\Lambda$.

\medskip
\noindent{\it Proof.}
In the half-plane $r(z)>0$ the function $\Psi_{x=0}$  has poles at the points
$z_m+nh+2m\omega_2, n=1,2,\ldots, \ m\in Z$. By definition,
the function $\Psi_{x=1}$ is holomorphic in $\Pi_{x=1}$. Therefore, in the strip
$\Pi_1$ the matrix $S$ has the poles at the points congruent to $z_m$
mod $\Lambda_h$. Then equations (\ref{mS}) imply (\ref{eS}).

We refer to above defined correspondence
\beq\label{emap}
\{\rho_i, A_m^{ij}, q_i\}\longmapsto \{S_0^i, S_m^{\,ij}, s_i\}
\eeq
as the direct monodromy map.

\subsection{Local monodromies}

All the results that were obtained above for the case of difference equations with
rational coefficients have analogs in the elliptic case. For example, the analogues of
the special regular singular equations are equations (\ref{el1}) with coefficients
$A(z)$ such that their residues $A_m$ of $A(z)$ are rank 1 matrices,
and the determinant of $A$ identically equals $1$, $\det\, A(z)=1,$
and such that the parameters $q_i$ in (\ref{elA}) satisfy the constraint
\beq\label{econst}
\sum_{i=1}^r q_i=0.
\eeq
The space of such matrices will be denoted by $\A_0(\G)$.
The dimension of $\A_0(\G)$ equals $\dim\, \A_0(\G)=n(2r-1)-n+(r-1)=(2n+1)(r-1)$.
The first term in the last equation is the dimension of the subspace of matrix
functions of the form (\ref{elA}) having rank 1 residues. The second term is
the number of conditions equivalent to the constraint $\det A=1$. The last term is
the number of parameters $q_i$.
Let $\B(\G)=\A_0(\G)/C^{r-1}$ be the quotient of $\A_0(\G)$ under the action
$A\to gAg^{-1}$, where $g$ is the diagonal matrix. The dimension of $\B(\G)$ equals
$\dim\, \B(\G)=2n(r-1)$. Explicit parameterization of an open set of the space $\B(\G)$
can be obtained as follows. Let us order the poles, and consider matrices $A(z)$
of the form
\beq\label{eA1}
A(z)=L_{n}(z)L_{n-1}(z)\cdots L_{1}(z)
\eeq
where
\beq\label{eA2}
L_{m}^{ij}=f_m^i\  {\s(z-q_{i,m+1}+q_{j,m}-z_m)\over\,\s(z-z_m)\
\s(q_{i,m+1}-q_{j,m})}, \
\eeq
and $q_{i,m}$ are complex numbers satisfying (\ref{econst}) and such that
$q_{i,n+1}=q_{i,1}$.

The residue of $L_m$ at $z_m$ has rank 1. Therefore, its determinant has at most
simple pole at $z_m$. The constraint (\ref{econst}) for $q_{i,m}$ implies that $\det L_m$
is an elliptic function. Therefore, it is constant.
The vector $f_m$ can be normalized by the condition $\det L_m(z)=\det L(0)=1$
\beq\label{ede}
\prod_{i=1}^r f_i^{-1}=\det\left[{\s(z_m+q_{i,m+1}-q_{j,m})\over\,\s(z_m)\
\s(q_{i,m+1}-q_{j,m})}\right].
\eeq
The number of parameters $(f_{i,m},q_{i,m})$ in (\ref{eA1}) satisfying
the constraints (\ref{econst}) and (\ref{ede}) equals the dimension
of  $\B(\G)$.

Let us assume, that the first coordinates $r_{m}=r(z_m)$ of the poles of $A$ in the basis
$2\omega_{\a}$ are distinct $r_{l}<r_{m},\ l<m$. Below we use the notations
$r_{0}=0, \ r_{n+1}=1$.

\begin{theo} For a generic matrix $A\in \A_0(\G)$
the  equation (\ref{el1}) has a unique set of  meromorphic solutions
$\Psi_k,\ k=0,\ldots,n$, which are holomorphic
in the strips $r_{k}<r(z)<r_{k+1}+r(h)$
and satisfy the relation
\beq\label{mm1}
\Psi_k(z+2\omega_2)=e^{\pi i\hat q/\omega_1}\Psi_k(z)e^{-2\pi i\hat s_{k}/h}, \
\hat s_k^{ij}=s_{i,k}\delta^{ij}\,,
\eeq
and such that the local connection matrices $M_k=\Psi_k^{-1}\Psi_{k-1}, \ k=1,\ldots, n,$
have the form
\beq\label{em25}
M_k=\a_{i,k}\  {\s_{h}(z-s_{i,k}+s_{j,k-1}-z_k)\over\s_{h}(z-z_m)
\s_{h}(s_{i,k}-s_{j,k-1})}, \
\eeq
where $s_{i,k}$ and $\a_{i,k}$ satisfy the relations
\beq\label{ede1}
\sum_{i=1}^r s_{i,k}=0,\ \
\prod_{i=1}^r \a_{i,k}^{-1}=\det\left[{\s_{h}(z_k+s_{i,k}-s_{j,k-1})\over\s_{h}(z_k)
\s_{h}(q_{i,k}-q_{j,k-1})}\right]\, .
\eeq
The map $\{f^i_m,q_{i,m}\}\longmapsto \{\a_k^i,s_{i,k}\}$
is a one-to-one correspondence of open sets of the varieties defined by the constraints
(\ref{econst}, \ref{ede}) and (\ref{ede1}), respectively.
\end{theo}
{\it Proof.} The existence of a meromorphic solution $\Psi_k'$, which is holomorphic
in the strip $r_{k}<r(z)<r_{k+1}+r(h)$ and satisfies the relation (\ref{mm1})
follows from the Lemma 5.1.  The matrix $M_k'=(\Psi'_k)^{-1}\Psi'_{k-1}$
has period $h$, i.e. $M_k'(z+h)=M'_k(z)$. From (\ref{mm1}) it follows
that
$$M_k'(z+2\omega_2)=e^{2\pi i\hat s_{k}/h}M'_k(z)e^{-2\pi i\hat s_{k-1}/h}.$$
In the strip $\Pi_{r_k+r(h)}$ it has simple poles at the point
$z_k$, where its residue has rank 1. Therefore, {\it a'priory} it can be
represented in the form
\beq\label{em251}
M_k'=\widetilde\a_{i,k}\b_{j,k}\  {\s_{h}(z-s_{i,k}+s_{j,k-1}-z_k)\over\s_{h}(z-z_k)
\s_{h}(s_{i,k}-s_{j,k-1})}, \
\eeq
The solutions $\Psi'_k$ are unique up to the transformation $\Psi'_k=\Psi_k F_k$,
where $F_k$ is a diagonal matrix $F_k^{i}\delta^{ij}$. If we set $F_{k-1}^j=\b_{j,k}$,
then the corresponding matrix $M_k=F_k^{-1}M_k'F_{k-1}$ has the form (\ref{em25}).
The constraint (\ref{ede1}) is equivalent to the equation $\det M_k=1$.

The proof of the last statement of the theorem is reduced to the
Riemann-Hilbert problem on a set of lines $r(z)=r_{m}+\e$. The solvability of
the corresponding problem for a generic set of matrices $M_k$ follows from
the Riemann-Roch theorem.

\noindent
{\it Remark.} Elliptic analog of the unitary equations considered in the Section 2
can be defined for the case of real elliptic curves. A generalization
of the corresponding results obtained above for the rational case is straightforward.

\subsection{Isomonodromy transformations.}

The characterization of equations (\ref{el1}) on $\G$ having the same monodromy data is
a straightforward generalization of the corresponding results in the rational case.

From (\ref{elmon}) it follows, that the determinant of $A\in \A(\G)$ is an elliptic function
\beq\label{edelta}
\det A(z)=D(z)=c{\prod_{\a=1}^N \s(z-\z_{\a})\over \prod_{k=1}^n\s(z-z_k)^{h_k}},\ \
\sum_{\a=1}^{N} \z_{\a}=\sum_{k=1}^n h_kz_k,\ \ N=\sum_k h_k.
\eeq
As before, we denote the subspace  of matrix functions having fixed determinant
$D(z)$ by $\A_D(\G)\subset \A(\G)$.
\begin{lem} (i) Two matrix functions $A(z)$ and $A'(z)$ of the form
(\ref{elA}) under the map (\ref{emap}) correspond to the same
connection matrix $S(z)$ if and only if they are related by the equation
\beq\label{eiR}
A'(z)=R(z+1)A(z)R^{-1}(z),
\eeq
where the matrix $R$ has the following monodromy properties
\beq\label{e20}
R(z+2\omega_1)=R(z),
\ R(z+2\omega_2)=e^{\pi i \hat q'/\omega_1}R(z)\, e^{-\pi i \hat q/\omega_1}.
\eeq

(ii) If the zeros $\z_{\a}$ are not congruent, i.e. $(\z_{\a}-\z_{\b})h^{-1}\notin Z$, then
the monodromy correspondence (\ref{emap}) restricted to $\A_D(\G)$ is injective.
\end{lem}
The proof of the lemma follows directly form the definition of $S(z)$ and the monodromy
properties of the canonical solutions of difference equations.

Let us call the two elliptic functions $D$ and $D'$ equivalent, if the sets of
their poles $z_i, z_i'$ and zeros $\z_{\a},\z_{\a}'$ are congruent $mod \ h$
to each other, i.e
$(z_i-z_i')h^{-1}\in Z, \ \ (\z_{\a}-\z_{\a}')h^{-1}\in Z$.

\begin{theo} For each pair of equivalent elliptic functions $D$ and $D'$ there exists
a unique isomonodromy transformation
\beq\label{Trans1}
T_D^{D'}(\G):\A_D(\G)\longmapsto \A_{D'}(\G)
\eeq
\end{theo}
{\it Proof.}
Let $A(z)\in \A_D$ be a matrix of the form (\ref{elA}).
An elementary isomonodromy transformation of the first type is defined by
a pair $z_m, \z_{\a}$ and the left eigenvector $v$ of $A_m=\res_{z_m} A$,
corresponding to a non-zero eigenvalue $\lambda$ (see (\ref{eigen})

Consider the matrix $R(z)$ such that the entries of the inverse matrix have
the form
\beq\label{ei20}
\left(R^{-1}\right)^{ij}=p^i
\ {\s(z-q_i+q'_j-\z_{\a})\over \s(z-\z_{\a})},
\eeq
where $p^i$ are coordinates of the the null-vector of $A(\z_{\a})$,
\beq
\label{ei21}
A(\z_{\a})p=0.
\eeq
The residue of $R^{-1}$ at $\z_{\a})$ has rank $1$. Therefore, the determinant of $R^{-1}$
has one simple pole at $\z_{\a}$. If
the parameters $q_i'$ satisfy the condition
\beq\label{e30}
\sum_{i=1}^r q'_i=\z_{\a}-z_m+\sum_{i=1}^m q_i,
\eeq
then $\det R^{-1}$ has a zero at $z_m$. In the general position
the parameters $q_j'$ are uniquely defined by (\ref{e30}) and the equation
\beq\label{e31}
vR^{-1}(z_m)=0.
\eeq
Equation (\ref{e31}) implies that the matrix $R$ has the form:
\beq\label{ei201}
R^{ij}=
v^j\ {\s(z-q'_i+q_j-z_{m})\over \s(z-z_{m})},
\eeq
Consider now the matrix $A'$ given by (\ref{eiR}). From (\ref{ei21}) it follows that
$A'$ is regular at $\z_{\a}$. The matrix $A'$ has a pole of rank 1
at $z_m-1$. The rank of its residue at $z_m$ equals the rank of the matrix
$A_mR^{-1}(z_m)$. The left null-space of the last matrix contains the null-space of $A_m$
and the vector $v$. Hence, the residue of
$A'$ has rank $h_m-1$. As in the rational case, further iterations give a matrix
$T_i^{\a_1,\ldots,\a_{h_i}}(A)$, which is regular at $z_m$
and has a pole of rank $h_m$ at $z_m-h$.

As follows from Lemma 5.3, the isomonodromy transformation $T_m^{\a_1,\ldots,\a_{h_m}}$
is uniquely defined by the choice of a pole $z_m$ and a subset of $h_m$ zeros
$\z_{\a_s}$ of $D$.

An elementary isomonodromy transformation of the second type is defined by a pair of
zeros $\z_{\a}$ and $\z_{\b}$ of $D$.
Let $v_{\a}$ and $v_{\b}$ be the corresponding null-vectors, i.e.
\beq\label{e32}
A(\z_{a})v_{\a}=0;\ \  v_{\b}^TA(\z_{\b})=0.
\eeq
Then the same arguments as above, show that there exists a unique, up to a constant factor,
matrix $R=R_{\a,\b}$ of the form
\beq\label{ei230}
R^{ij}_{\a,\b}=v_{\b}^j\ {\s(z-q^{\a,\b}_i+q_j-\z_{\b}-h)\over \s(z-\z_{\b}-h)},
\eeq
and such that
\beq\label{ei240}
\left(R^{-1}_{\a,\b}\right)^{ij}=
v_{\a}^i\ {\s_1(z-q_i+q^{\a,\b}_j-\z_{\a})\over \s(z-\z_{\a})},
\eeq
Equations (\ref{e32}) imply that the matrix
$T^{\a|\b}(A)=R_{\a,\b}^{-1}(z+h)A(z)R_{\a,\b}^{-1}(z)$
is regular and non-degenerate at $\z_{\a}$ and $\z_{\b}$. It has the same set of poles
as $A$. The zeros of its determinant are $\z_{\a}-h,\ \z_{\b}+h$ and $\z_{\g},
\ \g\neq \a,\b$.

The transformation $T_D^{D'}(\G)$ can be obtained as a composition of elementary
isomonodromy transformations. The theorem is thus proven.

\bigskip
\noindent
{\bf Isomonodromy deformations changing elliptic curves.} The isomonodromy
transformations $T_D^{D'}(\G)$ are analogs of the isomonodromy transformations
constructed in Section 3 for difference equations with  rational coefficients.
In the elliptic case there exist isomonodromy transformations
which have no analog in the rational case for the obvious reason: they
change the periods of the corresponding elliptic curves.

Our next goal is to define an elementary isomonodromy transformation of the third kind
which keeps the poles of $A$ and zeros of its determinant fixed.

\begin{lem} For a generic matrix function $A(z)$ of the form (\ref{elA}) there exists
a meromorphic matrix function $\R(z)$, which is holomorphic in the strip $\Pi_*:
0<r(z)<1+r(h)$ and satisfies the following monodromy relations
\beq\label{e40}
\R(z+2\omega_1+h)A(z)=\R(z), \ \ \
\R(z+2\omega_2)=e^{2\pi i \hat q'/(2\omega_1+h)}\R(z)e^{-\pi i\hat q/\omega_1},
\eeq
where $\hat q'$ is diagonal. The function $\R$ is unique up to the transformation
$\R'=F\R$, where $F\in GL_r$ is a diagonal matrix.
\end{lem}
The function $\R$ satisfying the relations (\ref{e40}) can be regarded as
the canonical Bloch solution of difference equation (\ref{er}). Its existence can be proved
along with the lines identical to that used in the proof of the Lemma 5.1.

Consider now the matrix function $A'=\R(z+h)A(z)\R^{-1}(z)$. From (\ref{e40} )
it follows that
\beq\label{e50}
A'(z+2\omega_1+h)=A'(z),\ \ A'(z+2\omega_1)=e^{2\pi i \hat q'/(2\omega_1+h)}A'(z)
e^{-2\pi i \hat q'/(2\omega_1+h)}.
\eeq
Suppose, that the matrix $A$ is holomorphic and invertible in the strip $\Pi_{x=0}$.
Then $A'$ in the fundamental parallelogram, corresponding to the elliptic curve
with periods $(2\omega_1+h, 2\omega_2$ has the same poles $z_m$ as $A$.
In this parallelogram the zeros $\z_{\a}$ of its determinant coincide with the zeros
of $\det A$.

\noindent
{\it Remark.} If the conditions $r(h)<r(z_m), \ r(h)<r(\z_a)$ are not satisfied, then
an extra pole (or zero of the determinant) of $A'$ in $\Pi_{x=1}$ is congruent $(mod h)$
to the pole (or zero of the determinant) of $A'$ in $\Pi_0$.

\begin{theo} If the matrix $A$ is invertible in $\Pi_0$, then the above defined
transformation $A'=\R(z+h)A(z)\R^{-1}(z)$ is isomonodromic.
\end{theo}
For the proof of the theorem it is enough to note, that under the assumption of the
theorem the canonical Bloch solution $\Psi_1$ of (\ref{el1}) is holomorphic
and invertible in the strip $\Pi_{1+r(h)}$. Therefore, the Bloch solutions of equation
(\ref{el1}) with the coefficient $A'$, which define the connection
matrix $S'$ are equal to
\beq\label{4e}
\Psi'_{x=0}=\R\Psi_0, \ \ \Psi'_{1+r(h)}=\R\Psi_1.
\eeq
Hence, $S'(z)=S(z)$.

\end{document}